\documentclass[twocolumn,journal]{IEEEtran}
\usepackage[T1]{fontenc}
\usepackage[latin9]{inputenc}
\usepackage{booktabs}
\usepackage{amsmath}
\usepackage{amsthm}
\usepackage{amssymb}
\usepackage{graphicx}
\usepackage[unicode=true,
 bookmarks=true,bookmarksnumbered=true,bookmarksopen=true,bookmarksopenlevel=1,
 breaklinks=false,pdfborder={0 0 0},pdfborderstyle={},backref=false,colorlinks=false]
 {hyperref}
\hypersetup{pdftitle={Your Title},
 pdfauthor={Your Name},
 pdfpagelayout=OneColumn, pdfnewwindow=true, pdfstartview=XYZ, plainpages=false}

\makeatletter

\providecommand{\tabularnewline}{\\}

\theoremstyle{plain}
\newtheorem{thm}{\protect\theoremname}
\theoremstyle{remark}
\newtheorem{rem}[thm]{\protect\remarkname}
\theoremstyle{plain}
\newtheorem{lem}[thm]{\protect\lemmaname}

\usepackage[caption=false,font=footnotesize]{subfig}

\usepackage{cite}
\usepackage{cases}
\usepackage{algorithm}
\usepackage{algpseudocode}

\usepackage{hyperref}

\@ifundefined{showcaptionsetup}{}{%
 \PassOptionsToPackage{caption=false}{subfig}}
\usepackage{subfig}
\makeatother

\providecommand{\lemmaname}{Lemma}
\providecommand{\remarkname}{Remark}
\providecommand{\theoremname}{Theorem}

\begin{document}
\title{Optimal Sensor Placement for Source Localization: A Unified ADMM Approach}
\author{Nitesh~Sahu, Linlong~Wu,~\IEEEmembership{Member,~IEEE,}~Prabhu~Babu,
Bhavani~Shankar~M.~R.,~\IEEEmembership{Senior Member,~IEEE,}~and~Bj\"{o}rn~Ottersten,~\IEEEmembership{Fellow,~IEEE}\thanks{Copyright (c) 2015 IEEE. Personal use of this material is permitted.
However, permission to use this material for any other purposes must
be obtained from the IEEE by sending a request to pubs-permissions@ieee.org.
The first two authors contributed equally to this work. \emph{Corresponding
author: Linlong Wu.}}\thanks{Linlong~Wu, Bhavani~Shankar M.R. and Bj\"{o}rn~Ottersten are with
the Interdisciplinary Centre for Security, Reliability and Trust (SnT),
University of Luxembourg, 1855 Luxembourg City, Luxembourg. E-mail:
\{linlong.wu, bhavani.shankar, bjorn.ottersten\}@uni.lu. Their work
is supported in part by ERC AGNOSTIC under grant EC/H2020/ERC2016ADG/742648
and in part by FNR CORE SPRINGER under grant C18/IS/12734677.}\thanks{Nitesh~Sahu and Prabhu~Babu are with the Centre for Applied Research
in Electronics (CARE), Indian Institute of Technology Delhi, New Delhi\textendash 110016,
India. E-mail: \{nitesh.sahu, prabhubabu\}@care.iitd.ac.in.}}
\maketitle
\begin{abstract}
Source localization plays a key role in many applications including
radar, wireless and underwater communications. Among various localization
methods, the most popular ones are Time-Of-Arrival (TOA), Time-Difference-Of-Arrival
(TDOA), Angle-Of-Arrival (AOA) and Received Signal Strength (RSS)
based. Since the Cram\'{e}r-Rao lower bounds (CRLB) of these methods
depend on the sensor geometry explicitly, sensor placement becomes
a crucial issue in source localization applications. In this paper,
we consider finding the optimal sensor placements for the TOA, TDOA,
AOA and RSS based localization scenarios. We first unify the three
localization models by a generalized problem formulation based on
the CRLB-related metric. Then a \emph{u}nified op\emph{t}imization
fra\emph{m}ework for \emph{o}ptimal \emph{s}ensor placemen\emph{t
}(UTMOST) is developed through the combination of the alternating
direction method of multipliers (ADMM) and majorization-minimization
(MM) techniques. Unlike the majority of the state-of-the-art works,
the proposed UTMOST neither approximates the design criterion nor
considers only uncorrelated noise in the measurements. It can readily
adapt to to different design criteria (i.e. A, D and E-optimality)
with slight modifications within the framework and yield the optimal
sensor placements correspondingly. Extensive numerical experiments
are performed to exhibit the efficacy and flexibility of the proposed
framework. 
\end{abstract}


\IEEEpeerreviewmaketitle{}

\section{Introduction}

\IEEEPARstart{E}{nvironment} sensing via wireless sensor networks
(WSNs) has been of significant research interest over the past decade,
and one of their key applications is the target/source localization
\cite{akyildiz2002survey,li2002detection}. From here on, we will
use the terms target and source interchangeably. Typically, in source
localization, given some potentially noisy measurements from the sensors,
the position of the source is estimated based on various approaches.
A variety of source localization techniques exist in the literature
depending on the type of information measured and the source position
recovery mechanism from the observed data. The commonly used approaches
are based on time-of-arrival (TOA) \cite{shen2012accurate}, angle-of-arrival
(AOA) \cite{zhu2008network}, time-difference-of-arrival (TDOA) \cite{huang2014tdoa},
received-signal-strength (RSS) \cite{weiss2003accuracy} and frequency-difference-of-arrival
(FDOA) \cite{ho2004accurate}. Apart from the employed estimation
approaches, the localization accuracy also depends on the target-sensor
geometry \cite{yoo2018crs,yoo2020analysis}. Specifically, the mean
squared error (MSE) in the estimation of the source position will
be a function of the sensor geometry. Therefore, optimal sensor placement
is a key problem in source localization applications.

However, owing to the highly nonlinear dependence of the MSE on the
geometry, it is rather challenging to arrive at the optimal sensor
placements based on the MSE analysis. A viable alternative approach
is to derive the Cramér-Rao lower bound (CRLB) for the source localization
model and then optimize the bound with respect to the positions of
the sensors. Indeed, various schemes of sensor placement have been
proposed in the literature based on optimization of the CRLB matrix
or the Fisher information matrix (FIM, i.e., the inverse of the CRLB
matrix). The commonly used CRLB-based optimization criteria are the
A-optimality (i.e. minimizing the determinant of the CRLB) and D-optimality
(i.e., minimizing the trace of the CRLB) \cite{ucinski2004optimal}.

For the two-dimensional (2D) case, the optimal sensor geometries for
the AOA based model was obtained by optimizing the D-optimality criterion
in \cite{douganccay2008optimal}, which are shown to be lying in a
equiangular configuration. In \cite{xu2017optimal_aoa}, the authors
have considered the problem of optimal sensor placement for AOA\textendash localization
in 3D space, they have derived optimal sensing schemes by optimizing
the $A-$optimal criterion. In \cite{xu2020optimalHybrid}, the authors
have designed optimal sensing direction by optimizing the $A-$optimal
design criterion for hybrid RSS, AOA and TOA localization problem.
The authors of \cite{hamdollahzadeh2019optimal} have proposed optimal
sensor placement schemes for AOA-based localization problem with distance
dependent noise model. In \cite{fang2018frame}, the problem of optimal
sensor placement for AOA model using frame theory was addressed. The
authors of \cite{zheng2020toward} addressed the problem of optimal
access point deployment (in the context of wifi-based localization)
for AOA localization model. In \cite{zhao2013optimal}, for the TDOA
measurement model, the optimal geometry of the sensors is derived
to correspond to vertices of a $m$ sided regular polygon, $m$ being
the number of sensors. The same approach was later extended to the
3D case as well, showing the optimal geometry as centered platonic
solids (tetrahedron, cube, etc) where source is present at the center
and sensors at the vertices \cite{yang2005cramer}. In \cite{meng2016optimal},
the optimal sensor geometries were derived for the TDOA model by employed
the A-optimality criterion and considering both the centralized and
decentralized pairing configuration which are distinguished by a common
reference sensor. Furthermore, for the three-dimensional (3D) case,
The authors in \cite{xu2019optima_toa,xu2017optimal_aoa,xu2019optimal_rss}
studied the optimal sensor placement strategies by optimizing the
A-optimality criterion for the TOA, AOA and RSS methodologies, in
which they approximated the A-optimality criterion via a general inequality
and minimized the approximated criterion with respect to the sensor
positions. In \cite{rui2014elliptic}, the optimal receiver position
was determined based on A-optimality for both synchronous and asynchronous
elliptical positioning (2D and 3D) that minimizes the localization
error. The author in \cite{nguyen2021optimal} proposed a framework
for the optimal sensor-target geometries for different types of sensor
network with Bayesian priors by taking into account the uncertainty
in the prior knowledge of the target position. In \cite{heydari2020optimal},
an optimal sensor placement strategy has been presented for received
signal strength difference (RSSD) based localization problem with
unknown transmitted power by maximizing the determinant of FIM. In
\cite{nguyen2016optimal}, the authors studied the problem of optimal
geometry analysis for TOA localization, where they employed the D-optimal
design criterion to arrive at optimal geometry. In \cite{nguyen2015optimal},
a D-optimal design based optimal sensor placement strategy was explored
for cognitive radar application.

Although much research has been conducted in this field, several gaps
are still observed from the literature as follows: 
\begin{itemize}
\item The majority of the works assume that the noise in the measurements
are uncorrelated, which would lead to a simplified design criterion.
Although it is not unnatural to assume uncorrelated noise in the measurement
models, under some circumstances, the noise in some measurements could
actually be correlated. For example, in ocean applications, the deployed
hydrophones will be influenced by the action of the same regional
swell or flows, which may cause the ADC saturation and bring the correlated
measurement errors \cite{robinson2014good}. Even with the strong
assumption that the sensor measurements are uncorrelated, in the case
of TDOA-based source localization, the measurement noise covariance
matrix will be correlated (see equation \eqref{eq:CovMat_R_tdoa}).
Therefore, one can not rule out the possibility of correlated measurement
noise and it has to be considered while designing sensor placement
strategies.
\item It has been observed in the literature that various specialized algorithms/methodologies
were developed to design optimal sensor placement strategies by optimizing
different design criterion. There is clearly a lack of a unified framework
encompassing all the design criteria (like A- and D-optimality criteria)
for the various source localization methodologies. Such a framework,
while being theoretically elegant, should also offer flexibility to
the designer to incorporate more application-dependent constraints
on the sensor locations.
\item The design based on the E-optimality (i.e., minimizing the largest
eigenvalue of the CRLB) \cite{ucinski2004optimal} usually behaves
substantially more reliably with respect to minimization of the variances
of the parameter estimates. However, there is barely any method on
designing optimal placement strategies by optimizing the E-optimality,
one reason for which could be the difficulty of solving the associated
optimization problem.
\end{itemize}
To address the aforementioned gaps in the literature, we propose a
unified optimization framework for optimal sensor placement design
in this work. The key contributions of our work are mainly as follows: 
\begin{itemize}
\item We have formulated a general sensor placement problem to cover various
commonly considered cases on this research topic. This general formulation
subsumes the TOA, TDOA, AOA or RSS based model under the A-, D- or
E-optimal design criterion. Therefore, by solving this general problem,
it is expected that a unified solving approach can be developed.
\item Based on the general problem formulation, a unified optimization-based
framework has been proposed, which is to solve simpler sub-problems
in an iterative manner. The A-, D- or E-optimality can be handled
under the same umbrella of this framework with minor changes in the
subproblems. Unlike the state-of-the-art approaches which involve
handling highly nonlinear trigonometric functions directly or approximating
them necessarily, our unified framework neither handles any trigonometric
functions nor invokes any approximation in the associated optimization
problem.
\item To the best of our knowledge, the E-optimal design criterion was never
considered in the literature for optimal sensor placement. Our unified
framework encompasses the E-optimal design criterion. Additionally,
in the data models of the three source localization methods, we do
not assume that the noise in the model to be necessarily uncorrelated.
Our unified framework can readily handle the case of correlated noise
in the model.
\item Extensive numerical simulations has been performed for designing optimal
sensor placement for all three (TOA, TDOA, RSS\footnote{We will see that the problem formulations for RSS and AOA are essentially
the same, and hence, we presents the simulations for RSS for illustration.}) source localization methods for three optimal design criteria (A-,
D- and E-optimal designs).
\end{itemize}
The rest of the paper is organized as follows. In Section \ref{sec:Problem-Formulation},
we describe the system models for the TOA, TDOA, AOA and RSS based
source localization, and then formulate a unified CRLB based problem
of optimal sensor placement. In Section \ref{sec:Proposed-Algorithms},
an optimization approach to optimal sensor placement is developed
under the A-, D- and E-optimality design criteria for all the TOA,
TDOA and RSS models. Section \ref{sec:Simulation-Result-Analysis}
demonstrates numerical results. Conclusions are drawn in Section \ref{sec:Conclusion}.

\emph{Notations}: $\mathbb{R}^{n}$ and $\mathbb{R}^{m\times n}$
denote the $n$-dimensional real-valued vector space and $m\times n$
real-valued matrix space, respectively. Scalars, vectors and matrices
are denoted by standard lowercase letter $a$, lower case boldface
letter $\mathbf{a}$ and upper case boldface letter $\mathbf{A}$,
respectively. $\mathbf{A}\succeq\mathbf{B}$ represents that $\mathbf{A}-\mathbf{B}$
is a positive semidefinite matrix. The subscripts $\left(.\right)^{T}$,
$\left(.\right)^{-1}$, $\left(.\right)^{\frac{1}{2}}$ denote the
transpose, inverse and square root of a matrix, respectively. $\text{Tr}\left(.\right)$,
$\lambda_{m}\left(.\right)$, $\det\left(.\right)$ and $\left\Vert .\right\Vert _{F}$
denote the trace, maximum eigenvalue, determinant and Frobenius norm
of a matrix, respectively. $\left\Vert .\right\Vert _{p}$ and $\left|.\right|$
represents for the $\ell_{p}$ norm of a vector and the absolute value
of a scalar, respectively. $\mathbb{E}\left[.\right]$, $\log_{10}\left(.\right)$,
$\ln\left(.\right)$, $\mathbf{I}_{m}$, $\boldsymbol{1}_{m}$ and
$\frac{d}{dx}$ denote the statistical expectation, base-10 logarithm,
natural logarithm, $m\times m$ identity matrix, $m\times1$ vector
with all elements equal to $1$, and differentiation with respect
to $x$, respectively.

\section{System Models and A Unified Problem Formulation\label{sec:Problem-Formulation}}

Consider the problem of locating a stationary target in the 3D space
using $m$ sensors with known locations. The sensor may be active
with transmitting signals and receiving the echos (e.g. radar and
sonar) or passive receiving the signal reflected or transmitted by
the target (e.g. hydrophone and microphone). The static target is
assumed to be located at an unknown coordinates $\mathbf{p}\in\mathbb{R}^{n}$,
and $m$ stationary sensors are located at known coordinates $\mathbf{r}_{i}\in\mathbb{R}^{n},\forall i=1,\ldots m$
with $m\geq n+1$. Depending on the nature of the target (i.e, active
or passive) and sensors (i.e. types of measurement), several localization
methods can be deployed. In this section, the TOA, TDOA and RSS based
localization cases will be considered. For each of them, the system
model is first introduced followed by the derived CRLB. Based on these
CRLBs, a general problem is formulated to unify the sensor placements
of different models.

\subsection{System Model for TOA-Based Localization}

Considering the scenario of a passive target and multiple active sensors
measuring the round-trip TOA, the noisy measurement at the $i$-th
sensor is modeled as

\begin{equation}
\widetilde{t}_{i}=\frac{2\left\Vert \mathbf{p}-\mathbf{r}_{i}\right\Vert _{2}}{c}+n_{i},\forall i=1,\ldots m,\label{eq:toa_e1}
\end{equation}
where $\left\Vert \mathbf{p}-\mathbf{r}_{i}\right\Vert _{2}$ is the
target range from the $i$-th sensor, $c$ denotes the wave propagation
speed in the medium, and $n_{i}$ represents the TOA measurement noise.
It is usually assumed that $n_{i}$ follows a Gaussian distribution
\cite{joshi2008sensor,so2011linear,chepuri2014sparsity} denoted by
$n_{i}\thicksim\mathcal{N}\left(0,\sigma_{i}^{2}\right)$. Additionally,
it is worth mentioning that the design under the Gaussian CRLB yields
the best performance in the worst case over a large class of distributions
\cite{stoica2011gaussian}, which further validates this Gaussian
assumption. 

After converting the TOA measurement to the corresponding distance
measurement, we have

\begin{equation}
s_{i}=2\left\Vert \mathbf{p}-\mathbf{r}_{i}\right\Vert _{2}+cn_{i},\forall i=1,\ldots m,
\end{equation}
where $s_{i}\triangleq c\widetilde{t}_{i}$, $cn_{i}\thicksim\mathcal{N}\left(0,c^{2}\sigma_{i}^{2}\right)$,
and $c$ is the speed of light.

Concatenating all measurements from the $m$ sensors together, we
have the following measurement model

\begin{equation}
\mathbf{s}=2\mathbf{g}\left(\mathbf{p}\right)+\boldsymbol{\eta}_{toa},\label{eq:TOA_model}
\end{equation}
where $\mathbf{s}=\left[s_{1},\ldots,s_{m}\right]^{T}$ denotes the
measurements from the $m$ sensors, $\mathbf{g}\left(\mathbf{p}\right)=\left[\left\Vert \mathbf{p}-\mathbf{r}_{1}\right\Vert _{2},\ldots,\left\Vert \mathbf{p}-\mathbf{r}_{m}\right\Vert _{2}\right]^{T}$
and $\boldsymbol{\eta}_{toa}=\left[cn_{1},\ldots,cn_{m}\right]^{T}\sim\mathcal{N}\left(\boldsymbol{0},\mathbf{R}_{toa}\right)$
with the covariance matrix $\mathbf{R}_{toa}$ assumed to be a general
positive definite matrix. Consequently, the joint density function
of the observation vector $\mathbf{s}$ is given by
\begin{equation}
\begin{aligned}p\left(\mathbf{s};\mathbf{p}\right)= & \frac{1}{\left(2\pi\right)^{\frac{m}{2}}\sqrt{\det\left(\mathbf{R}_{toa}\right)}}\\
 & \exp\left(-\frac{1}{2}\left(\mathbf{s}-2\mathbf{g}\left(\mathbf{p}\right)\right)^{T}\mathbf{R}_{toa}^{-1}\left(\mathbf{s}-2\mathbf{g}\left(\mathbf{p}\right)\right)\right).
\end{aligned}
\end{equation}
As stated in the introduction, under some circumstances, especially
when some of the sensors are of a similar nature, their inherent noise
would be correlated due to the similar mechanism and hardware implementation.
Thus, $\mathbf{R}_{toa}$ will not necessarily be a diagonal matrix
in general.

We denote an unbiased estimate of the true target location by $\hat{\mathbf{p}}$,
the covariance matrix of which satisfies the following well-known
inequality \cite{kay1993fundamentals}

\begin{equation}
\mathbb{E}\left[\left(\hat{\mathbf{p}}-\mathbf{p}\right)\left(\hat{\mathbf{p}}-\mathbf{p}\right)^{T}\right]\succeq\mathbf{C}\left(\mathbf{p}\right)=\mathbf{F}^{-1}\left(\mathbf{p}\right),
\end{equation}
where $\mathbf{C}\left(\mathbf{p}\right)$ is the CRLB matrix, and
$\mathbf{F}\left(\mathbf{p}\right)$ is the Fisher information matrix
(FIM) \cite{kay1993fundamentals} given by 

\begin{equation}
\mathbf{F}\left(\mathbf{p}\right)=\mathbb{E}\left(\left(\frac{\partial\ln p\left(\mathbf{s};\mathbf{p}\right)}{\partial\mathbf{p}}\right)\left(\frac{\partial\ln p\left(\mathbf{s};\mathbf{p}\right)}{\partial\mathbf{p}}\right)^{T}\right)\in\mathbb{R}^{n\times n}.\label{eq:toa_e2}
\end{equation}
According to \eqref{eq:toa_e2}, the FIM for the TOA based localization
can be expressed as

\begin{equation}
\mathbf{F}_{toa}\left(\mathbf{p}\right)=4\mathbf{H}^{T}\mathbf{R}_{toa}^{-1}\mathbf{H},\label{eq:toa_e3}
\end{equation}
and the CRLB matrix is thereby

\begin{equation}
\mathbf{C}_{toa}\left(\mathbf{p}\right)=\mathbf{F}_{toa}^{-1}\left(\mathbf{p}\right)=\frac{1}{4}\left(\mathbf{H}^{T}\mathbf{R}_{toa}^{-1}\mathbf{H}\right)^{-1},
\end{equation}
where 

\begin{equation}
\mathbf{H}\triangleq\left[\begin{array}{c}
\mathbf{h}_{1}^{T}\\
\vdots\\
\mathbf{h}_{m}^{T}
\end{array}\right]\triangleq\left[\begin{array}{c}
\frac{\left(\mathbf{p}-\mathbf{r}_{1}\right)^{T}}{\left\Vert \mathbf{p}-\mathbf{r}_{1}\right\Vert _{2}}\\
\vdots\\
\frac{\left(\mathbf{p}-\mathbf{r}_{m}\right)^{T}}{\left\Vert \mathbf{p}-\mathbf{r}_{m}\right\Vert _{2}}
\end{array}\right].\label{eq:toa_e11}
\end{equation}
The matrix $\mathbf{H}$ is referred to as the orientation matrix,
in which each $\mathbf{h}_{i}$ is a unit vector (i.e. $\mathbf{h}_{i}^{T}\mathbf{h}_{i}=1$)
defining the orientation of the $i$-th sensor with respect to the
target. Therefore, the CRLB essentially depends only on the orientation
of sensors with respect to the target, and the ranges between the
target and sensors will not affect it. In other words, designing the
sensor placement is equivalent to designing the orientation of all
the sensors. 
\begin{rem}
The above parameterization of CRLB matrix in term of the unit vectors
$\left\{ \mathbf{h}_{i}\right\} $ is very different from the common
parameterization approach seen in the literature. In the current literature,
the elements of $\left\{ \mathbf{h}_{i}\right\} $ vectors are usually
specified in term of ``azimuth'' and ``elevation'' angles, and
thus the resulting CRLB matrix will be a complicated function of trigonometric
functions. Differently, we prefer to keep the elements of $\left\{ \mathbf{h}_{i}\right\} $
in Cartesian coordinates (we can easily compute the corresponding
azimuth and elevation angle from $\mathbf{h}_{i}$), which enables
us to develop a neat optimization method later. Hence,the CRLB matrix
\eqref{eq:toa_e3} will be treated as a function of the orientation
matrix $\mathbf{H}$.
\end{rem}
\begin{rem}
The noise in the model (\ref{eq:TOA_model}) is assumed to be distance
independent (distance from the target). However, when the noise is
distance dependent, one can constrain that the optimal design on a
predefined sphere for 3D case or circle for 2D case, which would lead
to a similar constraint ($\mathbf{h}_{i}^{T}\mathbf{h}_{i}=c^{2}$,
where $c$ denotes the radius of the sphere or the circle). Moreover,
our approach in this work can also handle the case where $\mathbf{h}_{i}^{T}\mathbf{h}_{i}=c_{i}^{2}$
with arbitrarily predefined $c_{i}$ for each sensor, which can been
seen clearly in Algorithm \ref{alg:Alg_subproblem_H}.
\end{rem}

\subsection{System Model for TDOA-Based Localization}

Consider the target to be active and each sensor receives the wave
transmitted by the target. We assume that the sensors are ideally
time synchronized but unsynchronized with the target's clock. Upon
receiving the wave transmitted by the target, each sensor estimates
the TOA of the wave as 

\begin{equation}
\widetilde{t}_{i}=t_{0}+\frac{\left\Vert \mathbf{p}-\mathbf{r}_{i}\right\Vert _{2}}{c}+n_{i},\forall i=1,\ldots,m,
\end{equation}
where $t_{0}$ represents the unknown time at which target transmits
the wave, $n_{i}\thicksim\mathcal{N}\left(0,\sigma_{i}^{2}\right)$
denotes the measurement error of the TOA of the wave, and $\mathbf{p}$,
$\mathbf{r}_{i}$ and $c$ are the same as in the previous subsection.
Converting the time measurements to the distance measurements, we
get

\begin{equation}
s_{i}=\left\Vert \mathbf{p}-\mathbf{r}_{i}\right\Vert _{2}+cn_{i},\forall i=1,\ldots,m,
\end{equation}
where $s_{i}\triangleq c\left(\widetilde{t}_{i}-t_{0}\right)$ and
$cn_{i}\thicksim\mathcal{N}\left(0,c^{2}\sigma_{i}^{2}\right)$ . 

Since $t_{0}$ is unknown, TOA difference or range difference can
be used as an alternative for source localization. One of the sensors
is set as the reference or anchor sensor, with respect to which the
range difference or TOA difference is computed. Without loss of generality,
considering the first sensor as the reference sensor, we can compute
the range difference as follows:

\begin{equation}
s_{i1}=\left\Vert \mathbf{p}-\mathbf{r}_{i}\right\Vert _{2}-\left\Vert \mathbf{p}-\mathbf{r}_{1}\right\Vert _{2}+cn_{i}-cn_{1},\forall i=2,\ldots,m.\label{eq:tdoa_e1}
\end{equation}

Concatenating all $\left\{ s_{i1}\right\} $ in a vector form, we
have

\begin{equation}
\mathbf{s}=\mathbf{K}\mathbf{g}\left(\mathbf{p}\right)+\mathbf{K}\mathbf{n},\label{eq:tdoa_e2}
\end{equation}
where $\mathbf{s}=\left[s_{21},\ldots,s_{m1}\right]^{T}$, $\mathbf{n}=\left[cn_{1},\ldots,cn_{m}\right]^{T}$
and

\begin{equation}
\mathbf{K}\triangleq\left[\begin{array}{cc}
-\mathbf{1}_{m-1} & \mathbf{I}_{m-1}\end{array}\right]=\left[\begin{array}{ccccc}
-1 & 1 & 0 & \ldots & 0\\
-1 & 0 & 1 & \ldots & 0\\
\vdots & \vdots & \vdots & \ddots & \vdots\\
-1 & 0 & 0 & \ldots & 1
\end{array}\right].
\end{equation}
Denoting $\boldsymbol{\eta}_{tdoa}\triangleq\mathbf{K}\mathbf{n}$,
then \eqref{eq:tdoa_e2} becomes

\begin{equation}
\mathbf{s}=\mathbf{K}\mathbf{g}\left(\mathbf{p}\right)+\boldsymbol{\eta}_{tdoa},
\end{equation}
where $\boldsymbol{\eta}_{tdoa}\sim\mathcal{N}\left(\boldsymbol{0},\mathbf{R}_{tdoa}\right)$
with

\begin{equation}
\mathbf{R}_{tdoa}=\mathbb{E}\left[\boldsymbol{\eta}_{tdoa}\boldsymbol{\eta}_{tdoa}^{T}\right]=\mathbf{K}\mathbb{E}\left[\mathbf{n}\mathbf{n}^{T}\right]\mathbf{K}^{T}.\label{eq:CovMat_R_tdoa}
\end{equation}

\begin{rem}
From equation \eqref{eq:CovMat_R_tdoa}, it can be seen clearly that
$\mathbf{R}_{tdoa}$ will not be diagonal even if $\mathbb{E}\left[\mathbf{n}\mathbf{n}^{T}\right]$
is diagonal. Thus, in the case of the TDOA based localization, the
covariance matrix $\mathbf{R}_{tdoa}$ by nature is a non-diagonal
positive definite matrix.
\end{rem}
Similarly, the FIM and CRLB for the TDOA based localization can be
expressed as, respectively,

\begin{equation}
\mathbf{F}_{tdoa}\left(\mathbf{p}\right)=\mathbf{H}^{T}\mathbf{K}^{T}\mathbf{R}_{tdoa}^{-1}\mathbf{K}\mathbf{H}\label{eq:toa_e3-1}
\end{equation}
and

\begin{equation}
\mathbf{C}_{tdoa}\left(\mathbf{p}\right)=\mathbf{F}_{tdoa}^{-1}\left(\mathbf{p}\right)=\left(\mathbf{H}^{T}\mathbf{K}^{T}\mathbf{R}_{tdoa}^{-1}\mathbf{K}\mathbf{H}\right)^{-1},
\end{equation}
in which the CRLB matrix is also a function of the orientation matrix
$\mathbf{H}$. 

\subsection{System Model for RSS-Based Localization}

In the RSS based source localization, the target transmits some specific
signal which is received by each sensor. Each sensor receives the
signal and measures the RSS at its own location. In the absence of
disturbance, the average power received at the $i$-th receiver is
modeled as \cite{song1994automatic} 

\begin{equation}
P_{i}=\frac{K_{i}P_{t}}{\left\Vert \mathbf{p}-\mathbf{r}_{i}\right\Vert _{2}^{\alpha}},\forall i=1,\ldots,m,
\end{equation}
where $P_{i}$ and $P_{t}$ denote the receiving and transmitted power,
respectively, $K_{i}$ accounts for all other factors which affect
the received power, and $\alpha$ denotes the path loss constant ($\alpha=2$
in case of free space). It is assumed that $K_{i}$, $P_{t}$ and
$\alpha$ are known \emph{a priori} obtained through calibration campaign
\cite{patwari2003relative,tarrio2008new}.

Due to shadow fading, the RSS disturbance is assumed to be log-normally
distributed \cite{so2011linear,rappaport1996wireless}. Thus, the
measured RSS at the $i$-th sensor in decibel scale is modeled as

\begin{equation}
\begin{aligned}10\log_{10}P_{i}= & 10\log_{10}K_{i}+10\log_{10}P_{t}\\
 & -10\alpha\log_{10}\left\Vert \mathbf{p}-\mathbf{r}_{i}\right\Vert _{2}+w_{i},
\end{aligned}
\label{eq:rss_e1}
\end{equation}
where the measurement error $w_{i}$ is now Gaussian distributed.
For the notation simplicity, by converting the base-$10$ logarithm
in \eqref{eq:rss_e1} to natural logarithm, we have
\begin{equation}
\ln P_{i}=\ln K_{i}+\ln P_{t}-\alpha\ln\left\Vert \mathbf{p}-\mathbf{r}_{i}\right\Vert _{2}+\left(0.1\ln10\right)w_{i},
\end{equation}
which can be further rewritten as 

\begin{equation}
z_{i}=-\alpha\ln\left\Vert \mathbf{p}-\mathbf{r}_{i}\right\Vert _{2}+\eta_{i}^{rss}\label{eq:rss_e2}
\end{equation}
where $z_{i}=\ln P_{i}-\ln K_{i}-\ln P_{t}$, $\eta_{i}^{rss}=\left(0.1\ln10\right)w_{i}$,
and $\eta_{i}^{rss}\thicksim\mathcal{N}\left(0,\sigma_{i}^{2}\right)$.
Collecting all the measurements from the $m$ sensors, the vector
matrix form is

\begin{equation}
\mathbf{z}=-\alpha\boldsymbol{\varphi}\left(\mathbf{p}\right)+\boldsymbol{\eta}_{rss},\label{eq:rss_e3-1}
\end{equation}
where $\mathbf{z}=\left[z_{1},\ldots,z_{m}\right]^{T}$, $\boldsymbol{\eta}_{rss}=\left[\eta_{1}^{rss},\ldots,\eta_{m}^{rss}\right]^{T}\sim\mathcal{N}\left(\boldsymbol{0},\mathbf{R}_{rss}\right)$,
and $\boldsymbol{\varphi}\left(\mathbf{p}\right)=\left[\ln\left\Vert \mathbf{p}-\mathbf{r}_{1}\right\Vert _{2},\ldots,\ln\left\Vert \mathbf{p}-\mathbf{r}_{m}\right\Vert _{2}\right]^{T}$.

Similar to the TOA-based source localization, the FIM based on model
\eqref{eq:rss_e3-1} can be computed as

\begin{equation}
\mathbf{F}_{rss}\left(\mathbf{p}\right)=\alpha^{2}\mathbf{H}^{T}\mathbf{D}^{T}\mathbf{R}_{rss}^{-1}\mathbf{D}\mathbf{H},\label{eq:rss_e3}
\end{equation}
where $\mathbf{D}=\mathrm{diag}\left(\left\Vert \mathbf{p}-\mathbf{r}_{1}\right\Vert _{2},\ldots,\left\Vert \mathbf{p}-\mathbf{r}_{m}\right\Vert _{2}\right)^{-1}$
is referred to as the range matrix. Consequently, the CRLB matrix
is 
\begin{equation}
\mathbf{C}_{rss}\left(\mathbf{p}\right)=\mathbf{F}_{rss}^{-1}\left(\mathbf{p}\right)=\frac{1}{\alpha^{2}}\left(\mathbf{H}^{T}\mathbf{D}^{T}\mathbf{R}_{rss}^{-1}\mathbf{D}\mathbf{H}\right)^{-1}.
\end{equation}

\begin{rem}
Unlike the TOA and TDOA cases, the RSS based CRLB depends on both
the orientation matrix $\mathbf{H}$ and the range matrix $\mathbf{D}$.
Here, we would like to note that in optimal sensor-target geometry
analysis, one of the key underlying assumption is that an initial
estimate of the target position is known by some other means. Consequently,
the sensors can be placed optimally based on the initial estimate
of the target \cite{nguyen2021optimal}, which in turn can further
refine the estimate of the target position. Therefore, the range matrix
$\mathbf{D}$ is known coarsely from each sensor based on an initial
estimate of the target position, and\textbf{ }we are more interested
in determining the orientation matrix $\mathbf{H}$ with respect to
that initial target position. 
\end{rem}
\begin{rem}
\label{rem:Remark 5}The optimal sensor placement for AOA model can
be easily included in our framework, a brief explanation on the same
is as follows. From \cite{xu2020optimalHybrid} and \cite{xu2017optimal_aoa},
the FIM for AOA\textendash based model would be
\end{rem}
\begin{equation}
\mathbf{F}_{aoa}\left(\mathbf{p}\right)=\mathbf{H}^{T}\mathbf{D}^{T}\mathbf{R}_{aoa}^{-1}\mathbf{D}\mathbf{H},\label{eq:eq1}
\end{equation}
where 

\begin{equation}
\mathbf{H}=\left[\begin{array}{cc}
-\sin\theta_{1} & \cos\theta_{1}\\
\vdots & \vdots\\
-\sin\theta_{m} & \cos\theta_{m}
\end{array}\right]
\end{equation}
and 

\begin{equation}
\mathbf{D}=\left[\begin{array}{ccc}
1/d_{1} & \ldots & 0\\
\vdots & \ddots & \vdots\\
0 & \ldots & 1/d_{m}
\end{array}\right]
\end{equation}
where $\theta_{1},\ldots,\theta_{m}$ are the orientation angles of
the sensors with respect to the target and $d_{1},\ldots,d_{m}$ denote
the distance between the sensors to the target. It can be seen from
\eqref{eq:eq1} that the FIM for AOA has the same structure as the
FIM for RSS model (i.e. equation (\ref{eq:rss_e3})) except for the
differences in $\mathbf{H}$. In fact, we will have the similar expression
for the AOA in the 3D case, and the corresponding $\mathbf{H}$ for
AOA can also be reparameterized as matrix with unit norm rows as the
norms of the rows of $\mathbf{H}$ would be a constant. To conclude,
the unified framework proposed in our work can easily include AOA
localization model.

\subsection{A General CRLB-Based Problem Formulation }

Up to this point, we have derived the three CRLB matrices for the
TOA, TDOA, AOA and RSS based models, and they are listed as follows:
\begin{equation}
\begin{cases}
\text{TOA}: & \mathbf{C}_{toa}\left(\mathbf{p}\right)=\frac{1}{4}\left(\mathbf{H}^{T}\mathbf{R}_{toa}^{-1}\mathbf{H}\right)^{-1}\\
\text{TDOA}: & \mathbf{C}_{tdoa}\left(\mathbf{p}\right)=\left(\mathbf{H}^{T}\mathbf{K}^{T}\mathbf{R}_{tdoa}^{-1}\mathbf{K}\mathbf{H}\right)^{-1}\\
\text{RSS}: & \mathbf{C}_{rss}\left(\mathbf{p}\right)=\frac{1}{\alpha^{2}}\left(\mathbf{H}^{T}\mathbf{D}^{T}\mathbf{R}_{rss}^{-1}\mathbf{D}\mathbf{H}\right)^{-1}\\
\text{AOA}: & \mathbf{C}_{aoa}\left(\mathbf{p}\right)=\mathbf{H}^{T}\mathbf{D}^{T}\mathbf{R}_{aoa}^{-1}\mathbf{D}\mathbf{H}^{-1}.
\end{cases}\label{eq:3_CRLB}
\end{equation}
It is clear to see that all the three expressions are functions of
the orientation matrix $\mathbf{H}$ and share the same structure. 

Ignoring the constant scalars of $\mathbf{C}_{toa}\left(\mathbf{p}\right)$,
$\mathbf{C}_{tdoa}\left(\mathbf{p}\right)$, $\mathbf{C}_{rss}\left(\mathbf{p}\right)$
and $\mathbf{C}_{aoa}\left(\mathbf{p}\right)$, a unified expression
for all the three CRLBs can be defined as 
\begin{equation}
\mathbf{C}\left(\mathbf{H}\right)=\left(\mathbf{H}^{T}\boldsymbol{\Phi}^{T}\mathbf{R}^{-1}\boldsymbol{\Phi}\mathbf{H}\right)^{-1},
\end{equation}
which will be reduced to one of the expressions of \eqref{eq:3_CRLB}
when $\mathbf{R}$ and $\boldsymbol{\Phi}$ are specified. Thus, this
general expression makes it viable to unify different sensor placement
problems with just a single one elegantly. Hereafter, we refer $\mathbf{C}\left(\mathbf{H}\right)$
as a general CRLB matrix.

Since $\mathbf{C}\left(\mathbf{H}\right)$ is a matrix, some function
is required to convert the goodness of $\mathbf{C}\left(\mathbf{H}\right)$
into a scalar value, which will serve as the evaluation or optimization
metric. Actually, in the CRLB based design or optimization, there
are many choices for this required function. Among them, the A-, D-
and E-optimality\footnote{The A-, D-, and E-optimal designs refer to trace $\mathrm{Tr}\left(.\right)$,
determinant $\det\left(.\right)$, or maximum eigenvalue $\lambda_{max}\left(.\right)$
of the CRLB matrix, respectively.} are the most widely used metrics \cite{atkinson2007optimum,pronzato2013design}.
For the sake of notation simplicity, a general scalar-valued function
$f\left(\cdot\right)$ is used to represent these three optimalities,
which will be specified later when solving the relevant problem. 

Therefore, based on the expression $\mathbf{C}\left(\mathbf{H}\right)$,
we have a general problem formulation for sensor placement, i.e.,
\begin{equation}
\begin{aligned} & \underset{\mathbf{H}}{\text{minimize}} & \mathrm{} & f\left(\mathbf{C}\left(\mathbf{H}\right)\right)\\
 & \text{subject to} &  & \mathbf{h}_{i}^{T}\mathbf{h}_{i}=1,\forall i=1,\ldots,m.
\end{aligned}
\label{eq:sec3_e1}
\end{equation}
Before addressing this problem, several points which are worth to
illustrate as follows:
\begin{itemize}
\item Problem \eqref{eq:sec3_e1} is a unified formulation in two aspects:
First, $\mathbf{C}\left(\mathbf{H}\right)$ can be one of the CRLBs
shown in \eqref{eq:3_CRLB} for the TOA, TDOA, AOA and RSS based source
localization. Second, the general function $f\left(.\right)$ can
be trace $\mathrm{Tr}\left(.\right)$, determinant $\det\left(.\right)$,
or maximum eigenvalue $\lambda_{m}\left(.\right)$ for A-, D- and
E-optimal designs. Thus, any method solving problem \eqref{eq:sec3_e1}
will make itself a unified approach to cover a lots of common cases
in the context of sensor placement. 
\item Problem \eqref{eq:sec3_e1} is nonconvex in both the objective functions
and constraints, which is challenging to tackle in general. Further,
in the case of E-optimal design, it is non-differentiable in general
due to $f\left(.\right)=\lambda_{m}\left(.\right)$. Consequently,
to the best of our knowledge, most of the analytic approaches can
only handle some special cases (e.g. a diagonal $\mathbf{R}$ and
only A or D-optimality). For heuristic approaches, the computational
cost make themselves less appealing especially for a large-scale sensor
network. Accordingly, the optimization approach to this general problem
would become quite competitive. 
\end{itemize}

\section{A Unified Optimization Approach to Optimal Sensor Placement\label{sec:Proposed-Algorithms}}

\subsection{Reformulation and The ADMM Framework}

The ADMM method is a powerful optimization framework, which has been
successfully applied to many convex and nonconvex problems. To tackle
problem \eqref{eq:sec3_e1}, the ADMM framework is adopted. Let $\mathcal{D}=\left\{ \mathbf{H}\in\mathbb{R}^{m\times n}\mid\mathbf{h}_{i}^{T}\mathbf{h}_{i}=1,\forall i=1,\ldots,m\right\} $
and introduce an auxiliary variable $\mathbf{X}$ such that $\boldsymbol{\Phi}\mathbf{H}=\mathbf{X}$,
then we can rewrite \eqref{eq:sec3_e1} as

\begin{equation}
\begin{aligned} & \underset{\mathbf{H}\in\mathcal{D},\mathbf{X}}{\text{minimize}} & \mathrm{} & f\left(\left(\mathbf{X}^{T}\mathbf{R}^{-1}\mathbf{X}\right)^{-1}\right)\\
 & \text{subject to} &  & \boldsymbol{\Phi}\mathbf{H}=\mathbf{X}.
\end{aligned}
\label{eq:sec3_e2}
\end{equation}
Its augmented Lagrangian is formed as 

\begin{equation}
\begin{aligned}L_{\rho}\left(\mathbf{X},\mathbf{H},\mathbf{G}\right)= & f\left(\left(\mathbf{X}^{T}\mathbf{R}^{-1}\mathbf{X}\right)^{-1}\right)+\mathrm{Tr}\left(\mathbf{G}^{T}\left(\boldsymbol{\Phi}\mathbf{H}-\mathbf{X}\right)\right)\\
 & +\frac{\rho}{2}\left\Vert \boldsymbol{\Phi}\mathbf{H}-\mathbf{X}\right\Vert _{F}^{2},
\end{aligned}
\label{eq:sec3_e3}
\end{equation}
where $\mathbf{G}\in\mathbb{R}^{m\times n}$ is the Lagrangian multiplier,
$\rho>0$ is the augmented Lagrangian parameter \cite{boyd2011distributed}. 

The standard ADMM update rules for problem \eqref{eq:sec3_e2} are
\cite{boyd2011distributed}:

\begin{subnumcases} 		
	\mathbf{X}_{k+1}=\text{arg}\underset{\mathbf{X}}{\text{min }}L_{\rho}\left(\mathbf{X},\mathbf{H}_{k},\mathbf{G}_{k}\right)\label{admm_X} 		
	\\ 		
	\mathbf{H}_{k+1}=\text{arg}\underset{\mathbf{\mathbf{H}\in\mathcal{D}}}{\text{min }}L_{\rho}\left(\mathbf{X}_{k+1},\mathbf{H},\mathbf{G}_{k}\right)\label{admm_H} 		
	\\		\mathbf{G}_{k+1}=\mathbf{G}_{k}+\rho\left(\boldsymbol{\Phi}\mathbf{H}_{k+1}-\mathbf{X}_{k+1}\right),\label{admm_G}
\end{subnumcases}The two subproblems (\ref{admm_X}) and (\ref{admm_H}) will be solved
subsequently in the following subsections.

\subsection{Solving the Subproblem of $\mathbf{X}$}

Given $\mathbf{H}_{k}$ and $\mathbf{G}_{k}$ at the $k$-th iteration,
$L_{\rho}\left(\mathbf{X},\mathbf{H}_{k},\mathbf{G}_{k}\right)$ can
be expressed as

\begin{equation}
\footnotesize{\begin{aligned} & L_{\rho}\left(\mathbf{X},\mathbf{H}_{k},\mathbf{G}_{k}\right)\\
= & f\left(\left(\mathbf{X}^{T}\mathbf{R}^{-1}\mathbf{X}\right)^{-1}\right)+\mathrm{Tr}\left(\mathbf{G}_{k}^{T}\left(\boldsymbol{\Phi}\mathbf{H}_{k}-\mathbf{X}\right)\right)+\frac{\rho}{2}\left\Vert \boldsymbol{\Phi}\mathbf{H}_{k}-\mathbf{X}\right\Vert _{F}^{2}\\
= & f\left(\left(\mathbf{X}^{T}\mathbf{R}^{-1}\mathbf{X}\right)^{-1}\right)+\frac{\rho}{2}\mathrm{Tr}\left(\mathbf{X}^{T}\mathbf{X}\right)-\mathrm{Tr}\left(\mathbf{D}_{k}^{T}\mathbf{X}\right)+\beta_{k},
\end{aligned}
}\label{eq:sec3_e4}
\end{equation}
where $\beta_{k}=\mathrm{Tr}\left(\mathbf{G}_{k}^{T}\boldsymbol{\Phi}\mathbf{H}_{k}\right)+\frac{\rho}{2}\mathrm{Tr}\left(\mathbf{H}_{k}^{T}\boldsymbol{\Phi}^{T}\boldsymbol{\Phi}\mathbf{H}_{k}\right)$
and

\begin{equation}
\mathbf{D}_{k}=\mathbf{G}_{k}+\rho\boldsymbol{\Phi}\mathbf{H}_{k}.\label{eq:sec3_Dk}
\end{equation}

Let $\mathbf{R}^{-\frac{1}{2}}\mathbf{X}=\mathbf{Y}$, and then \eqref{eq:sec3_e4}
can be written as 
\begin{equation}
\begin{aligned} & L_{\rho}\left(\mathbf{Y},\mathbf{H}_{k},\mathbf{G}_{k}\right)\\
= & f\left(\left(\mathbf{Y}^{T}\mathbf{Y}\right)^{-1}\right)+\frac{\rho}{2}\mathrm{Tr}\left(\mathbf{Y}^{T}\mathbf{R}\mathbf{Y}\right)-\mathrm{Tr}\left(\mathbf{E}_{k}^{T}\mathbf{Y}\right)+\beta_{k}
\end{aligned}
\end{equation}
with $\mathbf{E}_{k}=\mathbf{R}^{1/2}\mathbf{D}_{k}.$ Thus, the optimization
problem w.r.t $\mathbf{X}$ is equivalent to 
\begin{equation}
\begin{aligned} & \underset{\mathbf{Y}}{\text{minimize}} & \mathrm{} & f\left(\left(\mathbf{Y}^{T}\mathbf{Y}\right)^{-1}\right)+\frac{\rho}{2}\mathrm{Tr}\left(\mathbf{Y}^{T}\mathbf{R}\mathbf{Y}\right)-\mathrm{Tr}\left(\mathbf{E}_{k}^{T}\mathbf{Y}\right),\end{aligned}
\label{eq:subproblem_Y}
\end{equation}
where the term $f\left(\left(\mathbf{Y}^{T}\mathbf{Y}\right)^{-1}\right)$
is nonconvex and may not lead to any closed form solution. For some
complicated optimization problems that cannot be handled by a single
optimization technique, it has been demonstrated that MM could be
incorporated to solve the subproblem \cite{wu2019sequence,wei2021sparse}.
A brief review of the MM algorithm is as follows: 

The MM (MM stands for ``majorize-minimize'' for minimization problem
or ``minorize-maximize'' for maximization problem) algorithm is
an iterative method to solve an optimization problem. It works by
creating a surrogate function for the original objective function
(at each iteration) which either majorizes or minorizes the original
objective and this surrogate function is optimized instead of the
original objective. Let $f\left(u\right)$ be the objective to be
minimized then at $\tau-$th iteration the surrogate function $g\left(u\mid u_{\tau}\right)$
is created such that $f\left(u_{\tau}\right)=g\left(u_{\tau}\mid u_{\tau}\right)$
and $f\left(u\right)\leq g\left(u\mid u_{\tau}\right)$ for all $u$.
The next iterate $u_{\tau+1}$ is computed as follows:

\begin{equation}
u_{\tau+1}=\text{arg}\underset{u}{\text{min}}\,g\left(u\mid u_{\tau}\right).
\end{equation}
The iterations of MM monotonically decrease the objective function
and more details on MM can be found in \cite{sun2016majorization}.

We will solve problem \eqref{eq:subproblem_Y} using the majorization-minimization
(MM) technique , which will lead to a double-loop algorithm finally. 

At the $\tau$-th iteration of MM, the global bound of the objective
function of problem \eqref{eq:subproblem_Y} should be constructed,
which is provided in the following lemma.
\begin{lem}
\label{lem:Lemma4}The objective function of problem \eqref{eq:subproblem_Y}
is upper bounded by 
\begin{equation}
\begin{aligned}g_{L}\left(\mathbf{Y}\right)= & f\left(\left(\mathbf{Y}^{T}\mathbf{Y}\right)^{-1}\right)+\frac{\rho}{2}\lambda_{m}\left(\mathbf{R}\right)\mathrm{Tr}\left(\mathbf{Y}^{T}\mathbf{Y}\right)\\
 & -\mathrm{Tr}\left(\mathbf{A}_{k,\tau}^{T}\mathbf{Y}\right)-\frac{\rho}{2}\mathrm{Tr}\left(\mathbf{Y}_{\tau}^{T}\widetilde{\mathbf{R}}\mathbf{Y}_{\tau}\right),
\end{aligned}
\end{equation}
where $\lambda_{m}\left(\mathbf{R}\right)$ is the maximum eigenvalue
of $\mathbf{R}$, 
\begin{equation}
\widetilde{\mathbf{R}}=\mathbf{R}-\lambda_{m}\left(\mathbf{R}\right)\mathbf{I}_{m},
\end{equation}
\begin{equation}
\mathbf{A}_{k,\tau}=\mathbf{E}_{k}-\rho\widetilde{\mathbf{R}}\mathbf{Y}_{\tau},\label{eq:sec3_Ak}
\end{equation}
and the equality holds when $\mathbf{Y}=\mathbf{Y}_{\tau}$.
\end{lem}
\begin{IEEEproof}
See Appendix \ref{subsec:AppenA}.
\end{IEEEproof}
Within the MM iterations, the next update $\mathbf{Y}_{\tau+1}$ is
computed by solving the following problem:

\begin{equation}
\footnotesize{\begin{aligned} & \underset{\mathbf{Y}}{\text{minimize}} & \mathrm{} & f\left(\left(\mathbf{Y}^{T}\mathbf{Y}\right)^{-1}\right)+\frac{\rho}{2}\lambda_{m}\left(\mathbf{R}\right)\mathrm{Tr}\left(\mathbf{Y}^{T}\mathbf{Y}\right)-\mathrm{Tr}\left(\mathbf{A}_{k,\tau}^{T}\mathbf{Y}\right).\end{aligned}
}\label{eq:sec3_e8}
\end{equation}
Let $\mathbf{Y}_{*}$ be a minimizer of problem \eqref{eq:subproblem_Y}
computed by the MM iterations, then we can compute $\mathbf{X}_{k+1}$
as

\begin{equation}
\mathbf{X}_{k+1}=\mathbf{R}^{\frac{1}{2}}\mathbf{Y}_{*}.
\end{equation}
Note that the form of update equation for $\mathbf{X}_{k+1}$ is the
same for all TOA, TDOA and RSS based methods irrespective of the choice
of $f\left(.\right)$. However, the solution step of the problem in
\eqref{eq:sec3_e8} would be dependent on the choice of $f\left(.\right)$.
In the following, we will discuss how to solve problem \eqref{eq:sec3_e8}
when $f\left(.\right)=\mathrm{Tr}\left(.\right)$ (for A-optimal design),
$f\left(.\right)=\text{logdet}\left(.\right)$ (for D-optimal design)
and $f\left(.\right)=\lambda_{m}\left(.\right)$ (for E-optimal design).

\subsubsection{Problem \eqref{eq:sec3_e8} for A-Optimal Design }

In the case of A-optimal design, we have $f\left(.\right)=\text{Tr}\left(.\right)$,
and problem \eqref{eq:sec3_e8} becomes

\begin{equation}
\footnotesize{\begin{aligned} & \underset{\mathbf{Y}}{\text{minimize}} & \mathrm{} & \mathrm{Tr}\left(\left(\mathbf{Y}^{T}\mathbf{Y}\right)^{-1}\right)+\frac{\rho}{2}\lambda_{m}\left(\mathbf{R}\right)\mathrm{Tr}\left(\mathbf{Y}^{T}\mathbf{Y}\right)-\mathrm{Tr}\left(\mathbf{A}_{k,\tau}^{T}\mathbf{Y}\right),\end{aligned}
}\label{eq:Aopt_e1}
\end{equation}
where the objective function is denoted by $g_{A}\left(\mathbf{Y}\right)$.
Let $\mathbf{A}_{k,\tau}=\mathbf{U}\boldsymbol{\Sigma}\mathbf{V}^{T}$
be the singular value decomposition (SVD) of $\mathbf{A}_{k,\tau}$,
and then the SVD of $\mathbf{Y}$ can be written as $\mathbf{Y}=\mathbf{U}\boldsymbol{\varLambda}\mathbf{V}^{T}$,
where the singular value matrix $\boldsymbol{\varLambda}$ is unknown.
Therefore, $g_{A}\left(\mathbf{Y}\right)$ in problem \eqref{eq:Aopt_e1}
can be seen as a function of $\boldsymbol{\varLambda}$ only and can
be written as
\begin{equation}
\footnotesize{g_{A}\left(\boldsymbol{\varLambda}\right)\triangleq\mathrm{Tr}\left(\left(\boldsymbol{\varLambda}^{T}\boldsymbol{\varLambda}\right)^{-1}\right)+\frac{\rho}{2}\lambda_{m}\left(\mathbf{R}\right)\mathrm{Tr}\left(\boldsymbol{\varLambda}^{T}\boldsymbol{\varLambda}\right)-\mathrm{Tr}\left(\boldsymbol{\Sigma}^{T}\boldsymbol{\varLambda}\right),}
\end{equation}
which can be further rewritten as 

\begin{equation}
g_{A}\left(\left\{ \gamma_{i}\right\} \right)=\sum_{i=1}^{n}\varphi_{i}^{A}\left(\gamma_{i}\right),\label{eq:Aopt_e3}
\end{equation}
where $\varphi_{i}^{A}\left(\gamma_{i}\right)\triangleq\gamma_{i}^{-2}+\frac{\rho}{2}\lambda_{m}\left(\mathbf{R}\right)\gamma_{i}^{2}-\sigma_{i}\gamma_{i}$,
and $\left\{ \gamma_{i}\right\} _{i=1}^{n}$ and $\left\{ \sigma_{i}\right\} _{i=1}^{n}$
are singular values of $\mathbf{Y}$ and $\mathbf{A}_{k,\tau}$, respectively.

Let $\left\{ \hat{\gamma}_{i}\right\} _{i=1}^{n}$ be the minimizer
of $g_{A}\left(\left\{ \gamma_{i}\right\} \right)$ in \eqref{eq:Aopt_e3},
then 

\begin{equation}
\hat{\gamma}_{i}=\text{arg}\underset{\gamma_{i}}{\text{min}}\,\varphi_{i}^{A}\left(\gamma_{i}\right).\label{eq:e71}
\end{equation}
The minimizer $\hat{\gamma}_{i}$ can be easily computed numerically
as one of the positive roots of the quartic equation 
\begin{equation}
\frac{d\varphi_{i}^{A}\left(\gamma_{i}\right)}{d\gamma_{i}}=-2\gamma_{i}^{-3}+\rho\lambda_{m}\left(\mathbf{R}\right)\gamma_{i}-\sigma_{i}=0,
\end{equation}
Hence, the solution of problem \eqref{eq:Aopt_e1} is

\begin{equation}
\mathbf{Y}_{\tau+1}=\mathbf{U}\hat{\boldsymbol{\varLambda}}\mathbf{V}^{T},
\end{equation}
where $\hat{\boldsymbol{\varLambda}}$ is the diagonal matrix with
the elements $\left\{ \hat{\gamma}_{i}\right\} _{i=1}^{n}$.

\subsubsection{Problem \eqref{eq:sec3_e8} for D-Optimal Design}

In the case of $D$-optimal design, problem \eqref{eq:sec3_e8} becomes
\begin{equation}
\footnotesize{\begin{aligned}\underset{\mathbf{Y}}{\text{minimize}} & \mathrm{} & \mathrm{logdet}\left(\left(\mathbf{Y}^{T}\mathbf{Y}\right)^{-1}\right)+\frac{\rho}{2}\lambda_{m}\left(\mathbf{R}\right)\mathrm{Tr}\left(\mathbf{Y}^{T}\mathbf{Y}\right)-\mathrm{Tr}\left(\mathbf{A}_{k,\tau}^{T}\mathbf{Y}\right),\end{aligned}
}\label{eq:Dopt_e1}
\end{equation}
where the objective function is denoted as $g_{D}\left(\mathbf{Y}\right)$.

Similar to the $A-$Optimal design case, we can write $g_{D}\left(\mathbf{Y}\right)$
in term of singular value matrix $\boldsymbol{\varLambda}$ as 
\begin{equation}
\footnotesize{g_{D}\left(\boldsymbol{\varLambda}\right)\triangleq\mathrm{logdet}\left(\left(\boldsymbol{\varLambda}^{T}\boldsymbol{\varLambda}\right)^{-1}\right)+\frac{\rho}{2}\lambda_{m}\left(\mathbf{R}\right)\mathrm{Tr}\left(\boldsymbol{\varLambda}^{T}\boldsymbol{\varLambda}\right)-\mathrm{Tr}\left(\boldsymbol{\Sigma}^{T}\boldsymbol{\varLambda}\right),}
\end{equation}
which can be further rewritten as 

\begin{equation}
g_{D}\left(\left\{ \gamma_{i}\right\} \right)=\sum_{i=1}^{n}\varphi_{i}^{D}\left(\gamma_{i}\right),
\end{equation}
where $\varphi_{i}^{D}\left(\gamma_{i}\right)\triangleq-2\log\left(\gamma_{i}\right)+0.5\rho\lambda_{m}\left(\mathbf{R}\right)\gamma_{i}^{2}-\sigma_{i}\gamma_{i}$.

Let $\left\{ \hat{\gamma}_{i}\right\} _{i=1}^{n}$ be the minimizer
of $g_{D}\left(\boldsymbol{\varLambda}\right)$, then 

\begin{equation}
\hat{\gamma}_{i}=\text{arg}\underset{\gamma_{i}}{\text{min}}\,\varphi_{i}^{D}\left(\gamma_{i}\right)
\end{equation}
and $\hat{\gamma}_{i}$ is one of the positive roots of the quadratic
equation 
\begin{equation}
\frac{d\varphi_{i}^{D}\left(\gamma_{i}\right)}{d\gamma_{i}}=\rho\lambda_{m}\left(\mathbf{R}\right)\gamma_{i}-\sigma_{i}-2\gamma_{i}^{-1}=0,
\end{equation}
which can be computed as

\begin{equation}
\hat{\gamma}_{i}=\frac{\sigma_{i}+\sqrt{\sigma_{i}^{2}+8\rho\lambda_{m}\left(\mathbf{R}\right)}}{2\rho\lambda_{m}\left(\mathbf{R}\right)}.
\end{equation}
Hence, the solution of problem \eqref{eq:Dopt_e1} can be written
as

\begin{equation}
\mathbf{Y}_{\tau+1}=\mathbf{U}\hat{\boldsymbol{\varLambda}}\mathbf{V}^{T},
\end{equation}
where matrix $\hat{\boldsymbol{\varLambda}}$ is a diagonal matrix
with the elements $\left\{ \hat{\gamma}_{i}\right\} _{i=1}^{n}$.

\subsubsection{Problem \eqref{eq:sec3_e8} for E-Optimal Design}

In the E-optimal design case, problem \eqref{eq:sec3_e8} becomes
\begin{equation}
\footnotesize{\begin{aligned} & \underset{\mathbf{Y}}{\text{minimize}} & \mathrm{} & \lambda_{m}\left(\left(\mathbf{Y}^{T}\mathbf{Y}\right)^{-1}\right)+\frac{\rho}{2}\lambda_{m}\left(\mathbf{R}\right)\mathrm{Tr}\left(\mathbf{Y}^{T}\mathbf{Y}\right)-\mathrm{Tr}\left(\mathbf{A}_{k,\tau}^{T}\mathbf{Y}\right),\end{aligned}
}\label{eq:Eopt_e1}
\end{equation}
where the objective function is denoted by $g_{E}\left(\mathbf{Y}\right)$.

Similar to the A and D optimal designs, we can write $g_{E}\left(\mathbf{Y}\right)$
in term of the singular value matrix $\boldsymbol{\varLambda}$ as
\begin{equation}
\footnotesize{g_{E}\left(\boldsymbol{\varLambda}\right)\triangleq\lambda_{m}\left(\left(\boldsymbol{\varLambda}^{T}\boldsymbol{\varLambda}\right)^{-1}\right)+\frac{\rho}{2}\lambda_{m}\left(\mathbf{R}\right)\mathrm{Tr}\left(\boldsymbol{\varLambda}^{T}\boldsymbol{\varLambda}\right)-\mathrm{Tr}\left(\boldsymbol{\Sigma}^{T}\boldsymbol{\varLambda}\right),}
\end{equation}
which can be further rewritten as
\begin{equation}
g_{L}\left(\left\{ \gamma_{i}\right\} \right)=\underset{1\leq i\leq n}{\text{max }}\left\{ \frac{1}{\gamma_{i}^{2}}\right\} +\frac{\rho}{2}\lambda_{m}\left(\mathbf{R}\right)\sum_{i=1}^{n}\gamma_{i}^{2}-\sum_{i=1}^{n}\sigma_{i}\gamma_{i}.
\end{equation}
Therefore, in case of the E-optimal design, we have to solve the following
minimax problem

\begin{equation}
\begin{aligned} & \underset{\gamma_{i}}{\text{minimize}} & \mathrm{} & \underset{1\leq i\leq n}{\text{max }}\left\{ \frac{1}{\gamma_{i}^{2}}\right\} +\frac{\rho}{2}\lambda_{m}\left(\mathbf{R}\right)\sum_{i=1}^{n}\gamma_{i}^{2}-\sum_{i=1}^{n}\sigma_{i}\gamma_{i}\\
 & \text{subject to} &  & \gamma_{i}\geq0,\forall i=1,\ldots,n.
\end{aligned}
\label{eq:Eopt_e2}
\end{equation}
Unlike the previous cases, we will not have any closed form solution
for the nonconvex problem \eqref{eq:Eopt_e2}. However, through the
variable transform $\gamma_{i}^{2}=\theta_{i}$, problem \eqref{eq:Eopt_e2}
be reformulated using the epigraph form as

\begin{equation}
\begin{aligned} & \underset{\theta_{i},t}{\text{minimize}} & \mathrm{} & t+\frac{\rho}{2}\lambda_{m}\left(\mathbf{R}\right)\sum_{i=1}^{n}\theta_{i}-\sum_{i=1}^{n}\sigma_{i}\sqrt{\theta_{i}}\\
 & \text{subject to} &  & \theta_{i}\geq0,\forall i=1,\ldots,n\\
 &  &  & \frac{1}{\theta_{i}}\leq t,\forall i=1,\ldots,n\\
 &  &  & t\ge0.
\end{aligned}
\label{eq:Eopt_e4}
\end{equation}
Problem \eqref{eq:Eopt_e4} is convex in $n+1$ (with $n\leq3$) scalar
variables. Thus, it can be solved efficiently by some solvers like
CVX \cite{grant2013cvx}. Let $\left\{ \hat{\theta_{i}}\right\} $
be the solution of problem \eqref{eq:Eopt_e4}, then $\left\{ \hat{\gamma_{i}}\right\} $
can be obtained using $\hat{\gamma_{i}}=\sqrt{\hat{\theta_{i}}}$.
Hence, $\mathbf{Y}_{\tau+1}=\mathbf{U}\hat{\boldsymbol{\varLambda}}\mathbf{V}^{T}$
where matrix $\hat{\boldsymbol{\varLambda}}$ is a diagonal matrix
with the elements $\left\{ \hat{\gamma}_{i}\right\} _{i=1}^{n}$.

So far, we have derived the solving methods to problem (\ref{admm_X})
for all the A-, D- and E-optimal design criteria. The proposed algorithms
are summarized in Algorithm \ref{alg:Alg_subproblem_X}.

\begin{algorithm}[t]		 	
	\caption{Proposed method to problem (\ref{admm_X})}	 	
	\label{alg:Alg_subproblem_X}	 	
	\begin{algorithmic}[1]
		\Require $m,n,\rho,\mathbf{R},\mathbf{H}_{k},\mathbf{G}_{k}$
		\Ensure $\mathbf{X}_{k+1}$
		\State{$\mathbf{E}_{k}=\mathbf{R}^{\frac{1}{2}}\left(\mathbf{G}_{k}+\rho\boldsymbol{\Phi}\mathbf{H}_{k}\right)$}
		\State{$\widetilde{\mathbf{R}}=\mathbf{R}-\lambda_{m}\left(\mathbf{R}\right)\mathbf{I}_{m}$}
		\State{$\mathbf{Y}_{\tau}=\mathbf{R}^{-\frac{1}{2}}\mathbf{X}_{k}$}
		\State{$\tau=0$}
		\Repeat
			\State{$\mathbf{A}_{k,\tau}=\mathbf{E}_{k}-\rho\widetilde{\mathbf{R}}\mathbf{Y}_{\tau}$}
			\State{Compute the SVD $\mathbf{A}_{k,\tau}=\mathbf{U}\text{diag}\left(\left\{ \sigma_{i}\right\} \right)\mathbf{V}^{T}$}
			\State{$\begin{cases}\text{(I) \textbf{A-optimality criterion}:}\\\begin{aligned}\begin{array}{ll} & \text{Solve }\frac{2}{\gamma_{i}^{3}}-\rho\lambda_{m}\left(\mathbf{R}\right)\gamma_{i}+\sigma_{i}=0\text{ to obtain }\hat{\gamma_{i}}\end{array}\end{aligned} & \begin{alignedat}{1}\end{alignedat}\\\text{(II) \textbf{D-optimality criterion}:}\\\begin{aligned}\begin{array}{ll} & \hat{\gamma}_{i}=\frac{\sigma_{i}+\sqrt{\sigma_{i}^{2}+8\rho\lambda_{m}\left(\mathbf{R}\right)}}{2\rho\lambda_{m}\left(\mathbf{R}\right)}\end{array}\end{aligned} & \begin{alignedat}{1}\end{alignedat}\\\text{(III) \textbf{E-optimality criterion}:}\\\begin{array}{ll} & \text{Solve problem (\ref{eq:Eopt_e4}) to obtian }\hat{\theta_{i}}\\ & \hat{\gamma_{i}}=\sqrt{\hat{\theta_{i}}}\end{array}\end{cases}$}
\State{$\hat{\boldsymbol{\varLambda}}=\text{diag}\left(\left[\hat{\gamma_{1}},\ldots,\hat{\gamma_{n}}\right]\right)$}
\State{$\mathbf{Y}_{\tau+1}=\mathbf{U}\hat{\boldsymbol{\varLambda}}\mathbf{V}^{T}$}					
\State{$\tau\leftarrow\tau+1$}
		\Until Convergence
		\State{$\mathbf{X}_{k+1}=\mathbf{R}^{\frac{1}{2}}\mathbf{Y}_{\tau}$}
	\end{algorithmic}	 
\end{algorithm}

\subsection{Solving the Subproblem of $\mathbf{H}$}

As in the next step of ADMM, we compute $\mathbf{H}_{k+1}$, for that
we write the expression for $L_{\rho}\left(\mathbf{X}_{k+1},\mathbf{H},\mathbf{G}_{k}\right)$
as

\begin{equation}
\begin{aligned} & L_{\rho}\left(\mathbf{X}_{k+1},\mathbf{H},\mathbf{G}_{k}\right)\\
= & f\left(\left(\mathbf{X}_{k+1}^{T}\mathbf{R}^{-1}\mathbf{X}_{k+1}\right)^{-1}\right)+\mathrm{Tr}\left(\mathbf{G}_{k}^{T}\left(\boldsymbol{\Phi}\mathbf{H}-\mathbf{X}_{k+1}\right)\right)\\
 & +\frac{\rho}{2}\mathrm{Tr}\left(\left(\boldsymbol{\Phi}\mathbf{H}-\mathbf{X}_{k+1}\right)^{T}\left(\boldsymbol{\Phi}\mathbf{H}-\mathbf{X}_{k+1}\right)\right)\\
= & \frac{\rho}{2}\mathrm{Tr}\left(\mathbf{H}^{T}\boldsymbol{\Phi}^{T}\boldsymbol{\Phi}\mathbf{H}\right)+\mathrm{Tr}\left(\mathbf{C}_{k+1}^{T}\boldsymbol{\Phi}\mathbf{H}\right)+\alpha_{k+1},
\end{aligned}
\label{eq:sec3_e9}
\end{equation}
where $\mathbf{C}_{k+1}=\mathbf{G}_{k}-\rho\mathbf{X}_{k+1}$ and
\begin{equation}
\begin{aligned}\alpha_{k+1}= & f\left(\left(\mathbf{X}_{k+1}^{T}\mathbf{R}^{-1}\mathbf{X}_{k+1}\right)^{-1}\right)\\
 & +\mathrm{Tr}\left(\frac{\rho}{2}\mathbf{X}_{k+1}^{T}\mathbf{X}_{k+1}-\mathbf{G}_{k}^{T}\mathbf{X}_{k+1}\right).
\end{aligned}
\label{eq:alpha}
\end{equation}
Minimizing \eqref{eq:sec3_e9} with respect to $\mathbf{H}$ depends
on the choice of $\boldsymbol{\Phi}$, which has different expressions
for different localization models. In the following, we will derive
the updates of $\mathbf{H}$ for all three models separately.

\subsubsection{Update of $\mathbf{H}$ for RSS/AOA}

For the RSS based method, we substitute $\boldsymbol{\Phi}=\mathbf{D}$
in \eqref{eq:sec3_e9} and obtain 

\begin{equation}
\begin{aligned} & L_{\rho}\left(\mathbf{X}_{k+1},\mathbf{H},\mathbf{G}_{k}\right)\\
= & \frac{\rho}{2}\mathrm{Tr}\left(\mathbf{H}^{T}\mathbf{D}^{T}\mathbf{D}\mathbf{H}\right)+\mathrm{Tr}\left(\mathbf{C}_{k+1}^{T}\mathbf{D}\mathbf{H}\right)+\alpha_{k+1},
\end{aligned}
\end{equation}
which can be further rewritten (noting that $\mathbf{D}$ is diagonal
matrix) as 

\begin{equation}
\begin{aligned} & L_{\rho}\left(\mathbf{X}_{k+1},\mathbf{H},\mathbf{G}_{k}\right)\\
= & \frac{\rho}{2}\mathrm{Tr}\left(\sum_{i=1}^{m}d_{i}^{2}\mathbf{h}_{i}\mathbf{h}_{i}^{T}\right)+\mathrm{Tr}\left(\sum_{i=1}^{m}d_{i}\mathbf{c}_{i}^{k+1}\mathbf{h}_{i}^{T}\right)+\alpha_{k+1},
\end{aligned}
\label{eq:sec3_eq10}
\end{equation}
where $\mathbf{h}_{i}$ and $\mathbf{c}_{i}^{k+1}$ are the $i$-th
column of $\mathbf{H}^{T}$ and $\left(\mathbf{C}_{k+1}\right)^{T}$,
respectively. Since $\mathbf{h}_{i}^{T}\mathbf{h}_{i}=1$, $L_{\rho}\left(\mathbf{X}_{k+1},\mathbf{H},\mathbf{G}_{k}\right)$
can be further written as

\begin{equation}
L_{\rho}\left(\mathbf{X}_{k+1},\mathbf{H},\mathbf{G}_{k}\right)=\frac{\rho}{2}\sum_{i=1}^{m}d_{i}^{2}+\sum_{i=1}^{m}d_{i}\mathbf{h}_{i}^{T}\mathbf{c}_{i}^{k+1}+\alpha_{k+1}.\label{eq:sec3_eq11}
\end{equation}
The objective $L_{\rho}\left(\mathbf{X}_{k+1},\mathbf{H},\mathbf{G}_{k}\right)$
is separable in $\mathbf{h}_{i}$. 

Therefore, the minimizer $\hat{\mathbf{h}}_{i}$ of \eqref{eq:sec3_eq11}
for $\mathbf{H}\in\mathcal{D}$ is given by

\begin{equation}
\hat{\mathbf{h}}_{i}=-\mathbf{c}_{i}^{k+1}/\left\Vert \mathbf{c}_{i}^{k+1}\right\Vert _{2}\label{eq:sec3_eq12}
\end{equation}
and $\mathbf{H}_{k+1}$ is computed as $\mathbf{H}_{k+1}^{rss}=\left[\hat{\mathbf{h}}_{1},\ldots,\hat{\mathbf{h}}_{m}\right]^{T}.$

\subsubsection{Update of $\mathbf{H}$ for TOA}

For the TOA based model, the expression for $\mathbf{H}_{k+1}$ can
be derived by following the same derivation of the above RSS case
by setting $\boldsymbol{\Phi}=\mathbf{I}_{m}$. It finally leads to
the same expression as given in \eqref{eq:sec3_eq12} and thus $\mathbf{H}_{k+1}^{toa}=\left[\hat{\mathbf{h}}_{1},\ldots,\hat{\mathbf{h}}_{m}\right]^{T}.$

\subsubsection{Update of $\mathbf{H}$ for TDOA}

For the TDOA based model, considering $\boldsymbol{\Phi}=\mathbf{K}$
and $\mathbf{R}=\mathbf{R}_{tdoa}$, we have 

\begin{equation}
\begin{aligned} & L_{\rho}\left(\mathbf{X}_{k+1},\mathbf{H},\mathbf{G}_{k}\right)\\
= & \frac{\rho}{2}\mathrm{Tr}\left(\mathbf{H}^{T}\mathbf{M}\mathbf{H}\right)+\mathrm{Tr}\left(\mathbf{C}_{k+1}^{T}\mathbf{K}\mathbf{H}\right)+\alpha_{k+1},
\end{aligned}
\label{eq:sec3_tdoa_e1}
\end{equation}
where $\mathbf{M}\triangleq\mathbf{K}^{T}\mathbf{K}\succeq0$, and
$\alpha_{k+1}$ is defined by \eqref{eq:alpha} with $\mathbf{R}=\mathbf{R}_{tdoa}$.
Minimizing \eqref{eq:sec3_tdoa_e1} with respect to $\mathbf{H}\in\mathcal{D}$
is not straightforward, so similar to the previous subsection, we
employ MM to minimize $L_{\rho}\left(\mathbf{X}_{k+1},\mathbf{H},\mathbf{G}_{k}\right)$.
The term $\mathrm{Tr}\left(\mathbf{H}^{T}\mathbf{M}\mathbf{H}\right)$
in \eqref{eq:sec3_tdoa_e1} is convex of $\mathbf{H}$, and we reformulate
it via concave function as follows:

\begin{equation}
\begin{aligned} & L_{\rho}\left(\mathbf{X}_{k+1},\mathbf{H},\mathbf{G}_{k}\right)\\
= & \frac{\rho}{2}\mathrm{Tr}\left(\mathbf{H}^{T}\mathbf{M}\mathbf{H}-\lambda_{m}\left(\mathbf{M}\right)\mathbf{H}^{T}\mathbf{H}+\lambda_{max}\left(\mathbf{M}\right)\mathbf{H}^{T}\mathbf{H}\right)\\
 & \mathrm{Tr}\left(\mathbf{C}_{k+1}^{T}\mathbf{K}\mathbf{H}\right)+\alpha_{k+1}\\
= & \frac{\rho}{2}\mathrm{Tr}\left(\mathbf{H}^{T}\widetilde{\mathbf{M}}\mathbf{H}\right)+\frac{\rho}{2}\lambda_{m}\left(\mathbf{M}\right)\mathrm{Tr}\left(\mathbf{H}^{T}\mathbf{H}\right)\\
 & +\mathrm{Tr}\left(\mathbf{C}_{k+1}^{T}\mathbf{K}\mathbf{H}\right)+\alpha_{k+1},
\end{aligned}
\label{eq:sec3_tdoa_e2}
\end{equation}
where $\widetilde{\mathbf{M}}=\mathbf{M}-\lambda_{m}\left(\mathbf{M}\right)\mathbf{I}_{m}\preceq0$.
Since $\mathbf{h}_{i}^{T}\mathbf{h}_{i}=1,\forall i$ and $\mathrm{Tr}\left(\mathbf{H}^{T}\mathbf{H}\right)=m$,
\eqref{eq:sec3_tdoa_e2} becomes

\begin{equation}
L_{\rho}\left(\mathbf{X}_{k+1},\mathbf{H},\mathbf{G}_{k}\right)=\frac{\rho}{2}\mathrm{Tr}\left(\mathbf{H}^{T}\widetilde{\mathbf{M}}\mathbf{H}\right)+\mathrm{Tr}\left(\mathbf{C}_{k+1}^{T}\mathbf{K}\mathbf{H}\right)+\mu_{k+1},\label{eq:sec3_e59}
\end{equation}
where $\mu_{k+1}=\alpha_{k+1}+\frac{\rho}{2}m\lambda_{m}\left(\mathbf{M}\right)$. 

Since $\mathrm{Tr}\left(\mathbf{H}^{T}\widetilde{\mathbf{M}}\mathbf{H}\right)$
is concave, its upper bound\footnote{Similar to the procedure we handled in the update of $\mathbf{X}$,
we find the global upper bound for the function in (\ref{eq:sec3_e59})
at some given $\mathbf{H}=\mathbf{H}_{t}$ and obtain the next iterate
by minimizing the global upper bound.} at any $\mathbf{H}=\mathbf{H}_{t}$ can be the first order Taylor
series expansion as follows:

\begin{equation}
\mathrm{Tr}\left(\mathbf{H}^{T}\widetilde{\mathbf{M}}\mathbf{H}\right)\leq2\mathrm{Tr}\left(\mathbf{H}_{t}^{T}\widetilde{\mathbf{M}}\mathbf{H}\right)-\mathrm{Tr}\left(\mathbf{H}_{t}^{T}\widetilde{\mathbf{M}}\mathbf{H}_{t}\right).\label{eq:sec3_tdoa_e3}
\end{equation}
Using \eqref{eq:sec3_tdoa_e3}, the upper bound of $L_{\rho}\left(\mathbf{X}_{k+1},\mathbf{H},\mathbf{G}_{k}\right)$
can be written as

\begin{equation}
L_{\rho}\left(\mathbf{X}_{k+1},\mathbf{H},\mathbf{G}_{k}\right)\leq\mathrm{Tr}\left(\left(\mathbf{B}_{k,t}\right)^{T}\mathbf{H}\right)+\xi_{k,t},
\end{equation}
where $\xi_{k,t}=\mu_{k+1}-\frac{\rho}{2}\mathrm{Tr}\left(\mathbf{H}_{t}^{T}\widetilde{\mathbf{M}}\mathbf{H}_{t}\right)$
and $\mathbf{B}_{k,t}=\rho\widetilde{\mathbf{M}}^{T}\mathbf{H}_{t}+\mathbf{K}^{T}\mathbf{C}_{k+1}$.
The next update $\mathbf{H}_{t+1}$ at the $t$-th iteration of MM
is computed by solving

\begin{equation}
\mathbf{H}_{t+1}=\text{arg}\underset{\mathbf{H}\in\mathcal{D}}{\text{min}}\mathrm{Tr}\left(\left(\mathbf{B}_{k,t}\right)^{T}\mathbf{H}\right)=\text{arg}\underset{\mathbf{H}\in\mathcal{D}}{\text{min}}\sum_{i=1}^{m}\mathbf{h}_{i}^{T}\mathbf{b}_{i}^{k,t},\label{eq:sec3_tdoa_e5}
\end{equation}
where $\mathbf{b}_{i}^{k,t}$ is the $i$-th row of $\mathbf{B}_{k,t}$. 

Thus, the minimizer $\hat{\mathbf{h}}_{i}$ of \eqref{eq:sec3_tdoa_e5}
is given by

\begin{equation}
\hat{\mathbf{h}}_{i,t}=-\mathbf{b}_{i}^{k,t}/\left\Vert \mathbf{b}_{i}^{k,t}\right\Vert _{2}\label{eq:sec3_tdoa_e7}
\end{equation}
and $\mathbf{H}_{t+1}$ is computed as $\mathbf{H}_{t+1}=\left[\hat{\mathbf{h}}_{1,t},\ldots,\hat{\mathbf{h}}_{m,t}\right]^{T}.$

Let $\mathbf{H}_{*}$ be the minimizer of \eqref{eq:sec3_tdoa_e1},
which would be obtained after the convergence of the MM loop. Consequently,
for the ADMM update of $\mathbf{H}$, we have $\mathbf{H}_{k+1}=\mathbf{H}_{*}$.

So far, we have derived the update rule of $\mathbf{H}$ in the ADMM
framework, which differs slightly depending on the localization models.
The complete description of the derived algorithms to problem (\ref{admm_H})
is given in Algorithm \ref{alg:Alg_subproblem_H}.

\begin{algorithm}[t]		 	
	\caption{Proposed method to problem (\ref{admm_H})}	 	
	\label{alg:Alg_subproblem_H}	 	
	\begin{algorithmic}[1]
		\Require $m,n,\rho,\mathbf{R},\mathbf{X}_{k+1},\mathbf{G}_{k}$
		\Ensure $\mathbf{H}_{k+1}$
		\State{$\mathbf{C}_{k+1}=\mathbf{G}_{k}-\rho\mathbf{X}_{k+1}$}
		\State{$\begin{cases}\text{(I) \textbf{TOA based model}:}\\\begin{aligned}\begin{array}{ll} & \boldsymbol{\Phi}=\mathbf{I}_{m}\\ & \hat{\mathbf{h}}_{i}=-\mathbf{c}_{i}^{k+1}/\left\Vert \mathbf{c}_{i}^{k+1}\right\Vert _{2}\end{array}\end{aligned} & \begin{alignedat}{1}\end{alignedat}\\\text{(II) \textbf{TDOA based model}:}\\\begin{aligned}\begin{array}{ll} & \boldsymbol{\Phi}=\mathbf{K}\\ & t=0\\ & \textbf{repeat}\\ & \quad\;\widetilde{\mathbf{M}}=\mathbf{M}-\lambda_{m}\left(\mathbf{M}\right)\mathbf{I}_{m}\\ & \quad\;\mathbf{B}_{k,t}=\rho\widetilde{\mathbf{M}}^{T}\mathbf{H}_{t}+\mathbf{K}^{T}\mathbf{C}_{k+1}\\ & \quad\;\hat{\mathbf{h}}_{i,t}=-\mathbf{b}_{i}^{k,t}/\left\Vert \mathbf{b}_{i}^{k,t}\right\Vert _{2}\\ & \quad\; t\leftarrow t+1\\ & \text{\textbf{until}}\text{ convergence}\end{array}\end{aligned} & \begin{alignedat}{1}\end{alignedat}\\\text{(III) \textbf{RSS/AOA based model}:}\\\begin{aligned}\begin{array}{ll} & \boldsymbol{\Phi}=\mathbf{D}\\ & \hat{\mathbf{h}}_{i}=-\mathbf{c}_{i}^{k+1}/\left\Vert \mathbf{c}_{i}^{k+1}\right\Vert _{2}\end{array}\end{aligned} & \begin{aligned}\end{aligned}\end{cases}$}
		\State{$\mathbf{H}_{k+1}=\left[\hat{\mathbf{h}}_{1},\ldots,\hat{\mathbf{h}}_{m}\right]^{T}$}
	\end{algorithmic}	 
\end{algorithm}

\subsection{Summary of the Unified Approach and Computational Complexity}

In Algorithm \ref{alg:PropAlg}, we summarize the proposed unified
solving approach to the general sensor placement problem \eqref{eq:sec3_e2},
where $k$ denotes the index for the ADMM iterations. The indices
$\tau$ and $t$ used in the Algorithms \ref{alg:Alg_subproblem_X}
and \ref{alg:Alg_subproblem_H} respectively should not be confused
with the ADMM iteration index $k$. The index $\tau$ denotes the
index for the MM iterations for solving problem \eqref{eq:sec3_e8},
and $t$ denotes the index for the MM iterations for computing the
update $\mathbf{H}_{k+1}$ in the case of TDOA. It is noted that the
proposed approach is quite general to cover many models under different
optimal design criteria, which can be seen clearly from Algorithm
\ref{alg:Alg_subproblem_X}-\ref{alg:PropAlg}.

The main computational burden in the proposed algorithm is the computation
of SVD of $\mathbf{A}_{k,\tau}$ in the loop to compute $\mathbf{X}_{k+1}$
which can be computed with the computation complexity of $\mathcal{O}\left(m^{2}n+mn^{2}\right)$.
Apart from the SVD, all other computations in both the inner loop
and outer loop are computationally simpler. The computation of $\mathbf{R}^{1/2}$,
$\mathbf{R}^{-1/2}$, and $\lambda_{m}\left(\mathbf{R}\right)$ are
not dependent on the iterations and can be accomplished once outside
of the outer loop.

For $A-$ and $D-$optimal designs, the cost function in \eqref{eq:sec3_e1}
is nonconvex but smooth (as the objectives are differentiable), and
in the case of E-optimal design, the objective \eqref{eq:sec3_e1}
is non-convex and non-smooth (non-differentiable). It is well known
that the convergence of nonconvex ADMM to a general nonconvex problem
is still an open question. However, the approaches mentioned in \cite{wang2019global,liu2019linearized,hong2016convergence}
can be adapted to prove the convergence of ADMM iterations to a KKT
point of the respective optimal design problems. Moreover, we always
observed numerically the algorithm to be converging in our simulation
studies. The MM iterations employed in the update of $\mathbf{X}$
and as well as $\mathbf{H}$ (for TDOA case) do converge and its proof
of convergence can be found in \cite{razaviyayn2013unified}. 

\begin{algorithm}[t]		 	
	\caption{\textbf{U}nified op\textbf{T}imization fra\textbf{M}ework for \textbf{O}ptimal \textbf{S}ensor placemen\textbf{T} (UTMOST)}	 	
	\label{alg:PropAlg}	 	
	\begin{algorithmic}[1]
		\Require $m,n,\rho,\mathbf{R}\in\left\{ \mathbf{R}_{toa},\mathbf{R}_{tdoa},\mathbf{R}_{rss}\right\}, \boldsymbol{\Phi}=\left\{ \mathbf{I}_{m},\mathbf{K},\mathbf{D}\right\}$
		\Ensure $\mathbf{H}$
		\State{Set $k=0$}
		\State{Initialize $\mathbf{H}_{k}\in\mathcal{D}$ and $\mathbf{G}_{k}$}
		\State{$\mathbf{X}_{k}=\boldsymbol{\Phi}\mathbf{H}_{k}$}
		\Repeat
			\State{Calculate $\mathbf{X}_{k+1}$ via Algorithm \ref{alg:Alg_subproblem_X}}
			\State{Calculate $\mathbf{H}_{k+1}$ via Algorithm \ref{alg:Alg_subproblem_H}}
			\State{Update $\mathbf{G}_{k+1}$ by (\ref{admm_G})}
		\Until convergence 
		\State{$\mathbf{H}_{k+1}=\left[\hat{\mathbf{h}}_{1},\ldots,\hat{\mathbf{h}}_{m}\right]^{T}$}
	\end{algorithmic}	 
\end{algorithm}

\section{Simulation Results\label{sec:Simulation-Result-Analysis}}

In this section, we discuss the simulation results considering various
localization methodologies and different optimal design criteria.

\subsection{Sanity and Convergence Check}

First of all, we perform the sanity check or show the correctness
of the proposed algorithmic framework for the special case, when matrix
$\mathbf{R}$ is diagonal with same diagonal entries and $\boldsymbol{\Phi}=\mathbf{I}_{m}$
(thus it becomes the TOA model), for which the analytical optimal
values are known in the literature \cite{xu2019optima_toa}. When
$\mathbf{R}=\upsilon^{2}\mathbf{I}_{m}$ and $\boldsymbol{\Phi}=\mathbf{I}_{m}$
then optimal solution of \eqref{eq:sec3_e1} for $A-$, $D-$ and
$E-$optimality criteria satisfies the following relation \cite{neering2008optimal}

\begin{equation}
\mathbf{H}_{*}^{T}\mathbf{H}_{*}=\frac{m}{3}\mathbf{I}_{n},\label{eq:SimRes_e1}
\end{equation}
where $\mathbf{H}_{*}$ denotes the optimal solution for the case
when $\mathbf{R}=\upsilon^{2}\mathbf{I}_{m}$.

The theoretical value of the objective function for $A-$, $D-$ and
$E-$optimality criteria, denoted by $f_{A}^{theo}$, $f_{D}^{theo}$
and $f_{E}^{theo}$, and their corresponding numerical value obtained
by the proposed algorithms denoted by $f_{A}^{algo}$, $f_{D}^{algo}$
and $f_{E}^{algo}$ for the case $\mathbf{R}=\upsilon^{2}\mathbf{I}_{m}$
with $\upsilon=1$ are listed in table \ref{tab:TheoObjNumeObj}.
By applying \eqref{eq:SimRes_e1}, we can compute $f_{A}^{theo}$,
$f_{D}^{theo}$ and $f_{E}^{theo}$ as follows:
\begin{equation}
\begin{cases}
f_{A}^{theo}=\frac{9\upsilon^{2}}{m}\\
f_{D}^{theo}=\log\left(\frac{27\upsilon^{2}}{m^{3}}\right)\\
f_{E}^{theo}=\frac{3\upsilon^{2}}{m}.
\end{cases}
\end{equation}

\begin{table}[t]
\caption{Comparison between the theoretical and numerical optimal objective
values for the TOA based model \label{tab:TheoObjNumeObj}}

\centering{}%
\begin{tabular}{ccccccc}
\toprule 
$m$ & $f_{A}^{theo}$ & $f_{A}^{algo}$ & $f_{D}^{theo}$ & $f_{D}^{algo}$ & $f_{E}^{theo}$ & $f_{E}^{algo}$\tabularnewline
\midrule
\midrule 
$5$ & $1.8000$ & $1.8000$ & $-1.5324$ & $-1.5324$ & $0.60000$ & $0.60033$\tabularnewline
\midrule 
$10$ & $0.9000$ & $0.9000$ & $-3.6119$ & $-3.6119$ & $0.30000$ & $0.30004$\tabularnewline
\midrule 
$15$ & $0.6000$ & $0.6000$ & $-4.8283$ & $-4.8283$ & $0.20000$ & $0.20017$\tabularnewline
\midrule 
$20$ & $0.4500$ & $0.4500$ & $-5.6913$ & $-5.6913$ & $0.15000$ & $0.15003$\tabularnewline
\midrule 
$25$ & $0.3600$ & $0.3600$ & $-6.3607$ & $-6.3607$ & $0.12000$ & $0.12001$\tabularnewline
\bottomrule
\end{tabular}
\end{table}

From table \ref{tab:TheoObjNumeObj} it is observed that the optimal
value of the objective functions computed from the proposed algorithm
converge to their corresponding analytical value, hence confirming
the correctness of the proposed algorithmic framework.

\subsection{TOA Based Source Localization}

In this subsection, the proposed algorithmic framework is applied
to determine the optimal configuration of sensors around the target
to optimize the localization accuracy for TOA-based model. Without
loss of generality, we assume the target to be roughly located at
the origin, that is, $\mathbf{p}=\left[0,0,0\right]^{T}$. As illustrated
previously, the CRLB for the TOA-based model is independent of the
sensor-target distance, so the sensors are assumed to be on the unit
sphere and only their optimal orientations are to be determined. 

We take $m=6$, $n=3$ and the noise covariance matrix $\mathbf{R}_{toa}$
to be a general positive definite matrix given by

\begin{equation}
\mathbf{R}_{toa}=\left[\begin{array}{cccccc}
4.88 & 3.07 & -1.73 & 1.90 & 2.63 & -1.61\\
3.07 & 11.72 & -3.51 & 4.48 & 3.95 & 0.24\\
-1.73 & -3.51 & 21.82 & -1.20 & 0.49 & -4.74\\
1.90 & 4.48 & -1.20 & 3.63 & 3.71 & 1.00\\
2.63 & 3.95 & 0.49 & 3.71 & 8.45 & 0.56\\
-1.61 & 0.24 & -4.74 & 1.00 & 0.56 & 4.22
\end{array}\right].
\end{equation}
We initialize the proposed algorithm such that the sensors are uniformly
placed with respect to target with the following initialization

\begin{equation}
\mathbf{H}_{0}=\left[\begin{array}{ccc}
1 & 0 & 0\\
0 & 1 & 0\\
0 & 0 & 1\\
-1 & 0 & 0\\
0 & -1 & 0\\
0 & 0 & -1
\end{array}\right].\label{eq:SimRslt_e2}
\end{equation}
With this initialization we can compare the gain in accuracy with
the placement obtained by the proposed algorithm with respect to the
considered uniform placement and moreover we can also observe how
sensors change their positions with the iterations of the algorithm
from initial uniform placement to achieve a final optimal configuration.

\begin{figure}[t]
\begin{centering}
\subfloat[A-optimal design.]{\begin{centering}
\includegraphics[scale=0.25]{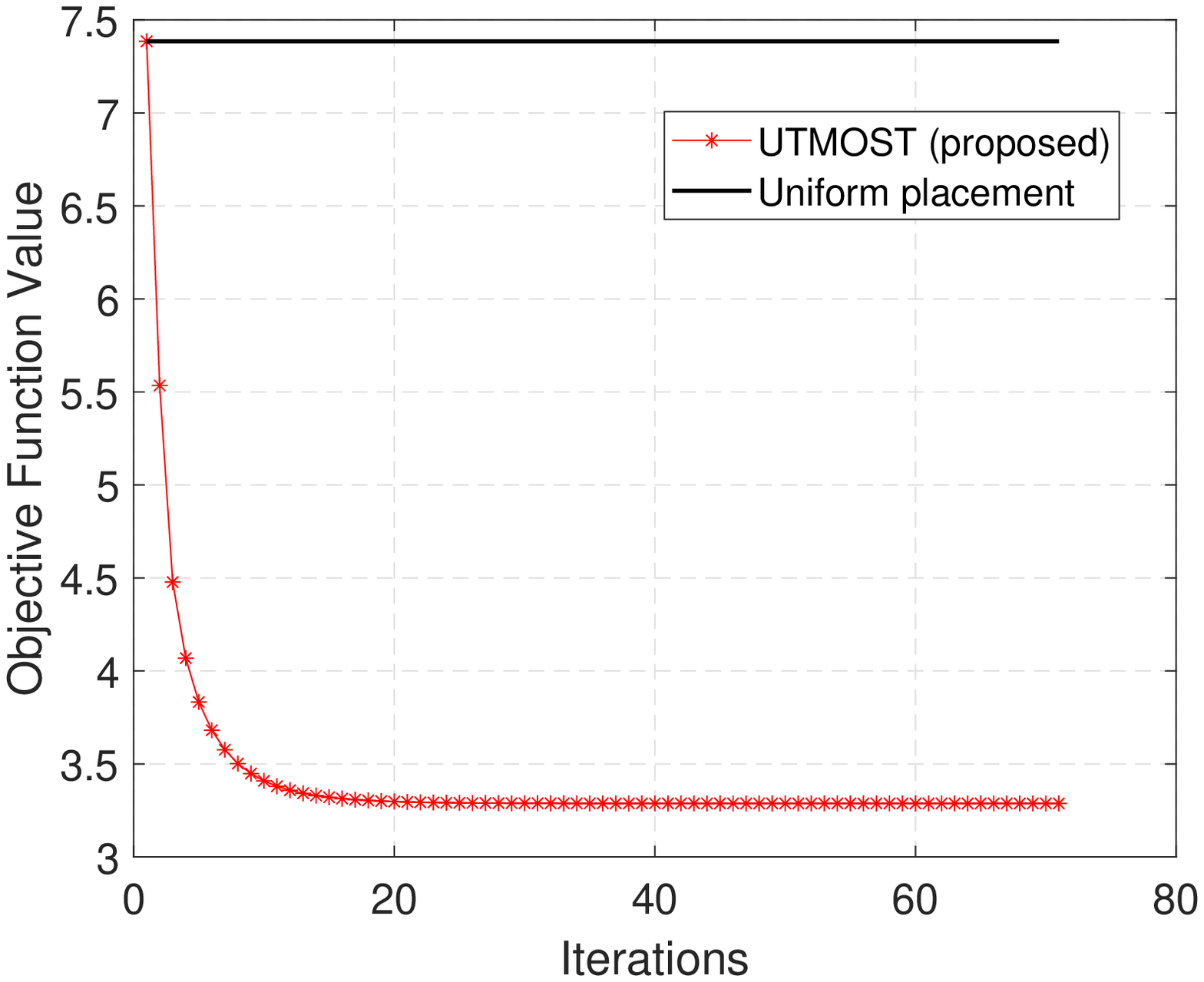}\includegraphics[scale=0.25]{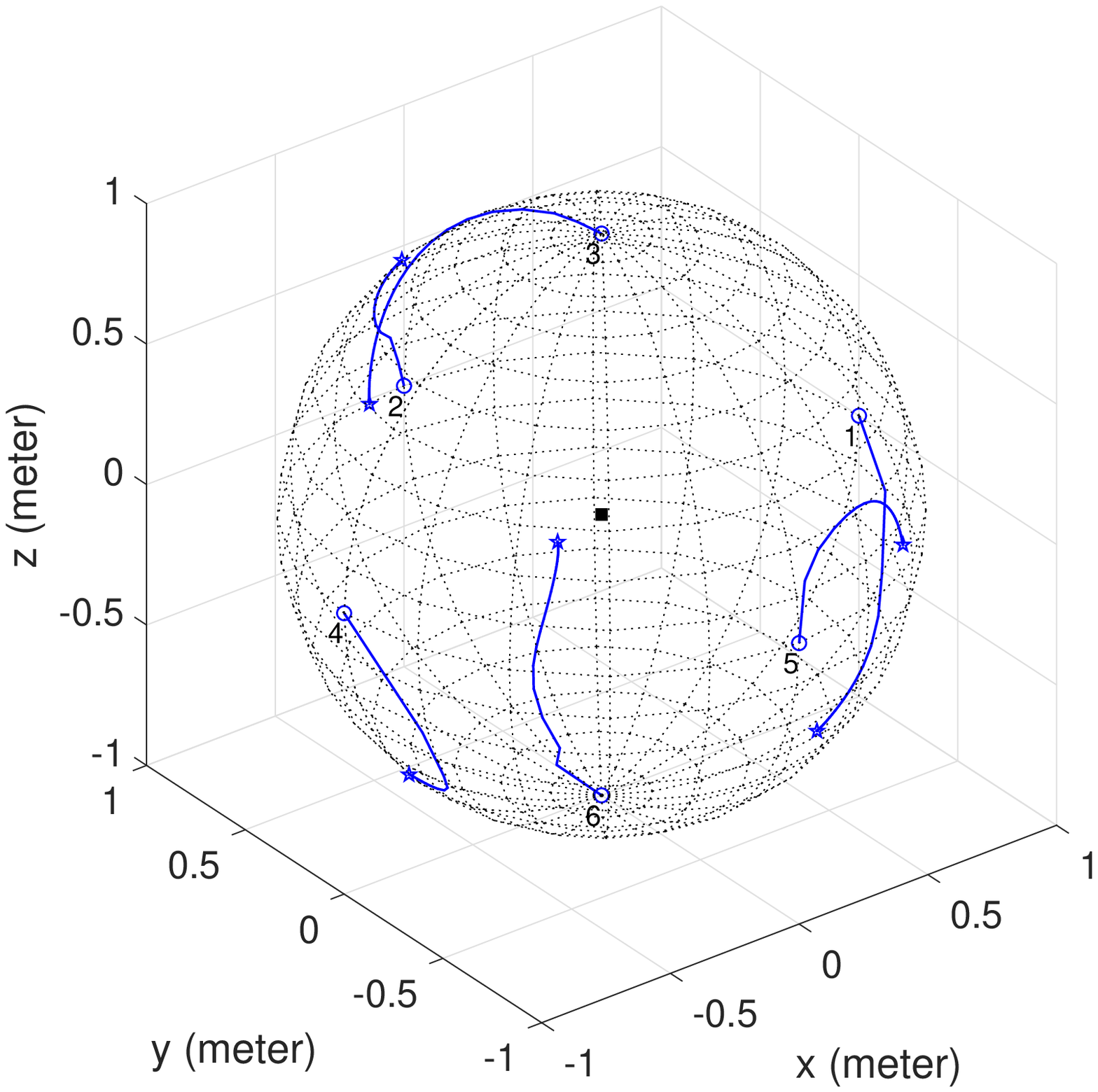}
\par\end{centering}
}
\par\end{centering}
\begin{centering}
\subfloat[D-optimal design]{\begin{centering}
\includegraphics[scale=0.25]{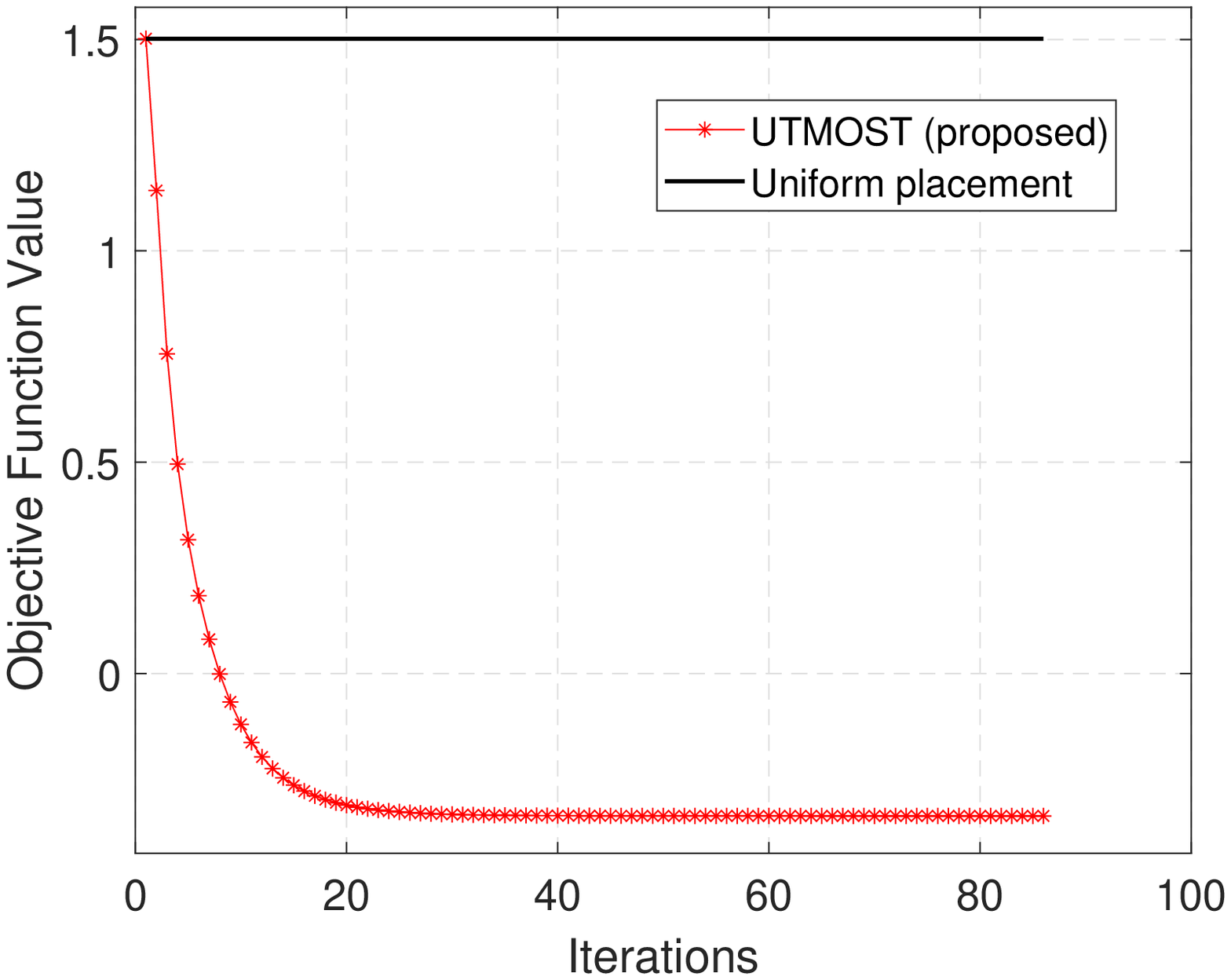}\includegraphics[scale=0.25]{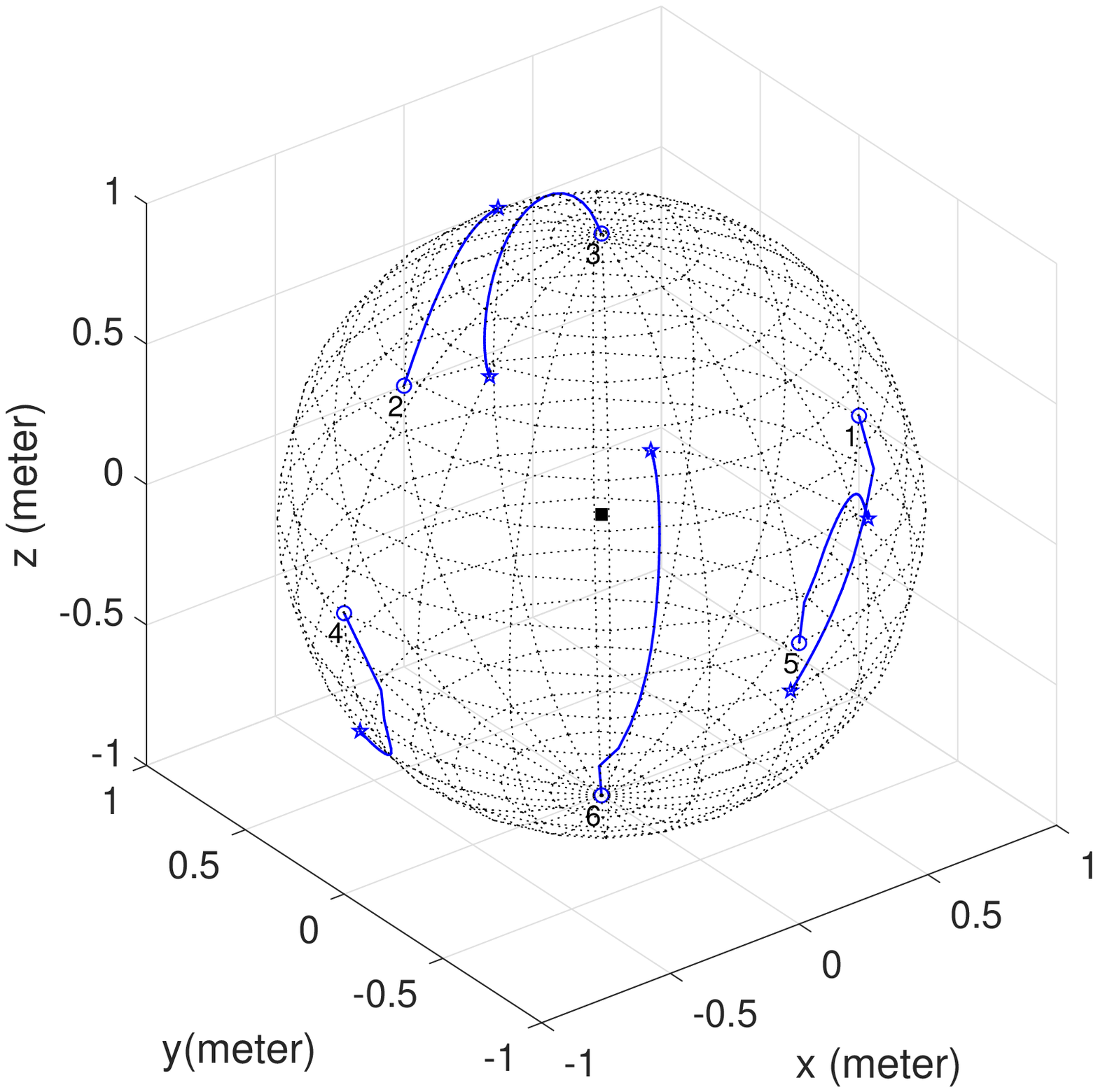}
\par\end{centering}
}
\par\end{centering}
\begin{centering}
\subfloat[E-optimal design.]{\begin{centering}
\includegraphics[scale=0.25]{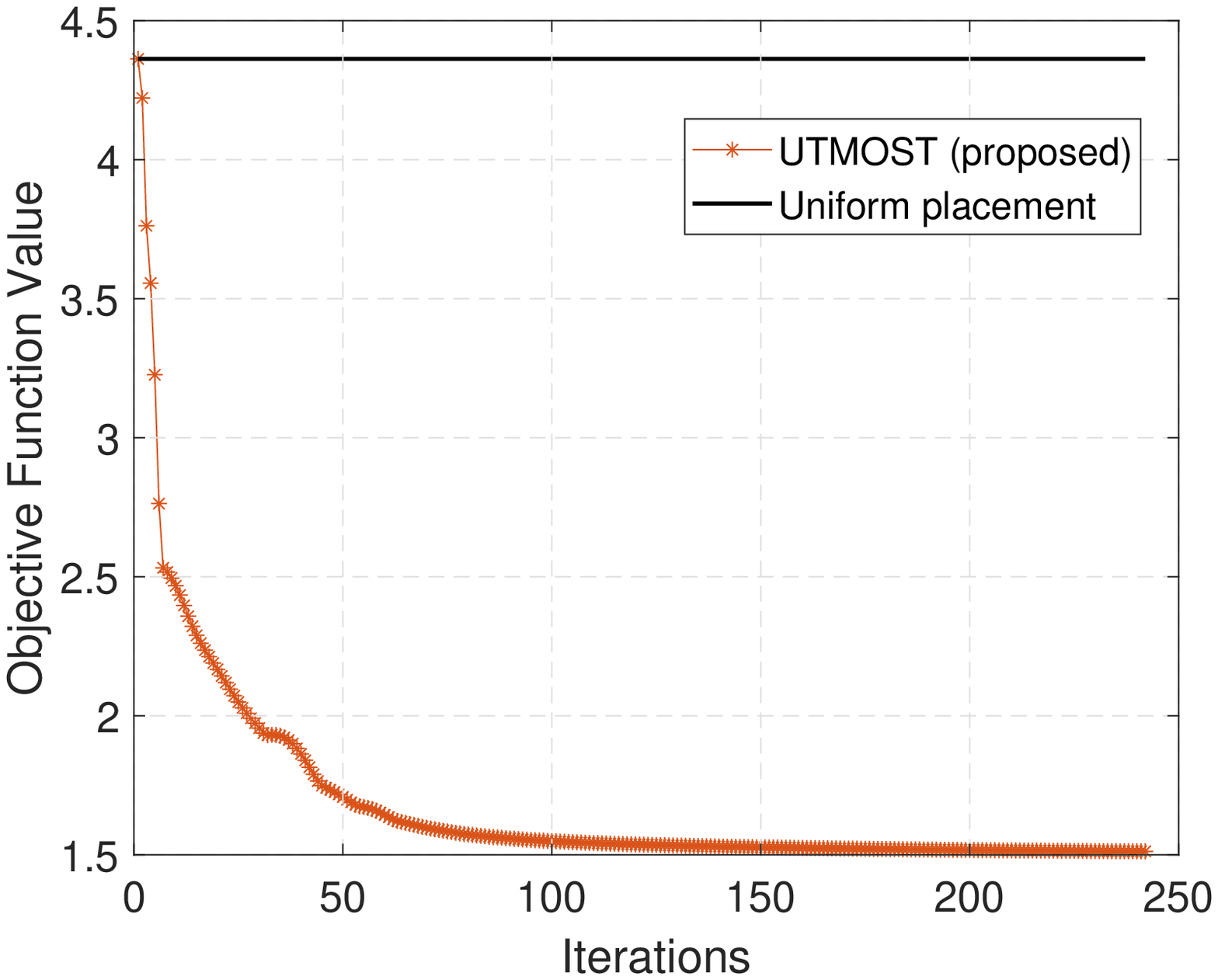}\includegraphics[scale=0.25]{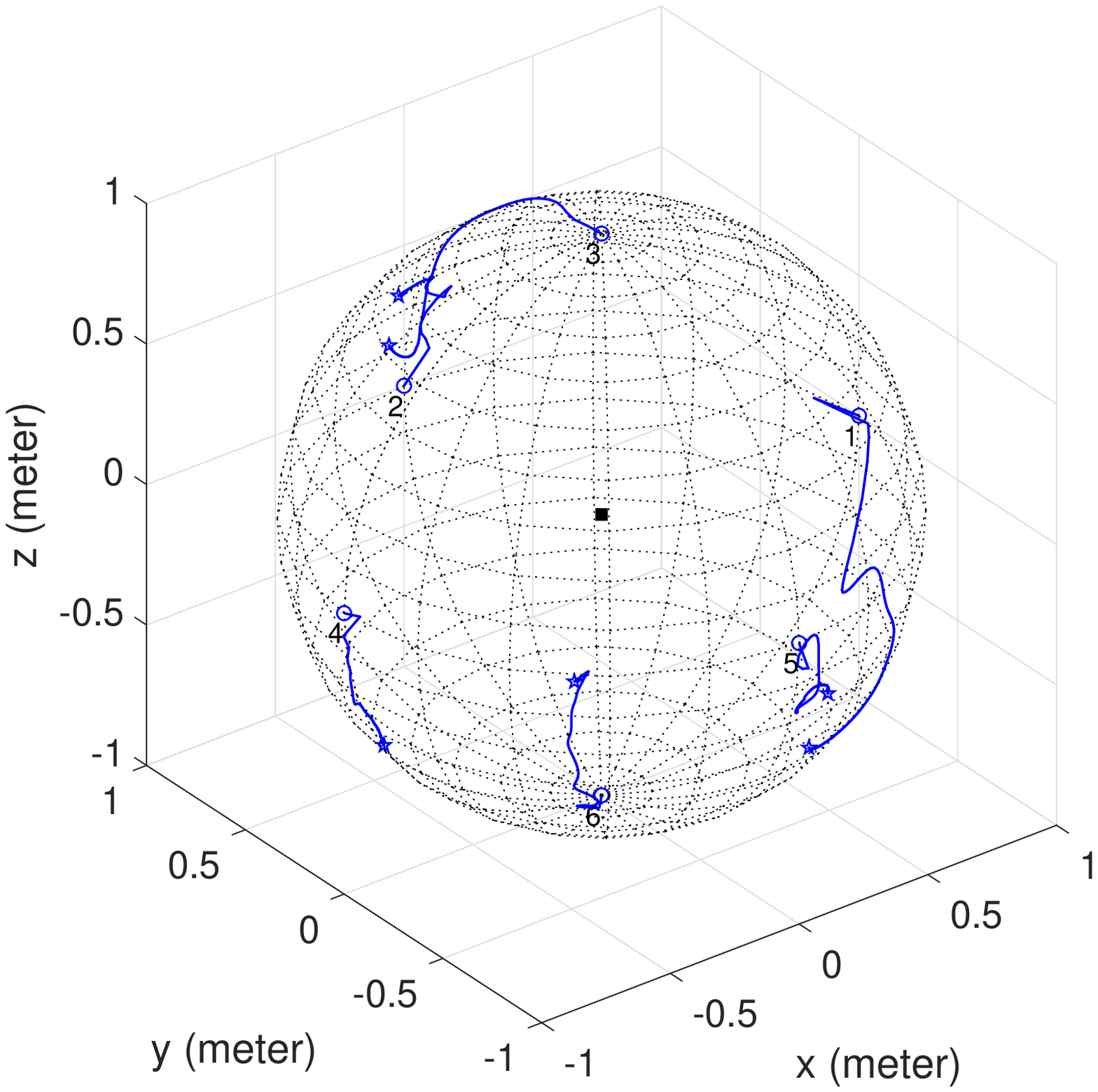}
\par\end{centering}
}
\par\end{centering}
\caption{\label{fig:ConvPlot_toa}Convergence plots and corresponding sensor
placements for the TOA based model under the correlated measurement
noise. Left: convergence plots; right: 3D view of sensor placement.
Black square: target; blue circle: initial position; blue pentagram:
end position.}
\end{figure}

In Figure \ref{fig:ConvPlot_toa}, we demonstrate the convergence
plots and the corresponding 3D placement trajectories of the proposed
method for all A-, D- and E-optimal designs. The baseline is the uniform
placement (i.e. the sensors are uniformly placed w.r.t. the target),
which is also set as the initial point of our proposed algorithm.
First, it is clear to see that from the convergence plots, our algorithm
monotonically decreases the design objective. Second, with reference
to the uniform placement, the proposed method (after convergence)
shows $55-70\%$ improvement in terms of the design criteria, which
implies further an enhancement of localization accuracy enabled by
the proposed algorithm. Therefore, to obtain the maximum localization
accuracy especially when the measurement noise is correlated, the
sensors must be placed in their corresponding optimal configuration,
which can be computed from the proposed algorithm.

Now, we apply a different sensor geometry in a 2D target localization
problem to investigate the estimation improvement by designing the
sensor placement. Specifically, we perform the maximum likelihood
estimation (MLE) in a 2D TOA based source localization problem under
different sensor placements, which includes random placement, uniform
placement, and the placement via our proposed algorithm for the A-optimal
design. To perform the simulation, the sensors are assumed to be located
on the circumference of a circle of unit radius whose center is at
origin and the target is assumed to be located at $\left(0.1,-0.3\right)$.
The MLE is implemented by conducting the 2D grid search to arrive
at the most probable target location and then performing the Gauss-Newton
algorithm \cite{dogancay2005instrumental}. The simulated noisy TOA
measurements (both uncorrelated and correlated noise case have been
considered) have been generated for uniform, optimal and for any randomly
selected placement. The results under different noises is provided
in Table \ref{tab:MLE-performance.}, where the MSE and bias of the
MLE are obtained using $1000$ Monte Carlo simulations. We can see
that the MSE and the bias of the estimates in case of optimal placement
are smaller than the uniform and random placements, which is consistent
to the theoretical conclusion that MLE asymptotically reach the trace
of CRLB \cite{xu2019optima_toa}. 

\begin{table}[t]
\caption{Comparison of the MLE performance for different placement\label{tab:MLE-performance.}}

\centering{}%
\begin{tabular}{cccc}
\toprule 
No. of sensors & Placement & MSE ($\text{m}^{2}$) & Bias ($\text{m}$)\tabularnewline
\midrule
\midrule 
 & Random & $0.2171$ & $0.1105$\tabularnewline
$m=3$ & Uniform & $0.1643$ & $0.0630$\tabularnewline
(Uncorrelated Noise) & Optimal & $0.1456$ & $0.0397$\tabularnewline
\midrule 
 & Random & $0.2070$ & $0.0998$\tabularnewline
$m=4$ & Uniform & $0.1260$ & $0.0231$\tabularnewline
(Uncorrelated Noise) & Optimal & $0.1156$ & $0.0169$\tabularnewline
\midrule 
 & Random & $31.0428$ & $1.1099$\tabularnewline
$m=5$ & Uniform & $9.2915$ & $1.5758$\tabularnewline
(Correlated Noise) & Optimal & $3.7194$ & $0.9597$\tabularnewline
\midrule 
 & Random & $0.9840$ & $0.3654$\tabularnewline
$m=7$ & Uniform & $1.2771$ & $0.6482$\tabularnewline
(Correlated Noise) & Optimal & $0.4711$ & $0.2483$\tabularnewline
\bottomrule
\end{tabular}
\end{table}

\begin{figure}[t]
\begin{centering}
\subfloat[\label{fig:SurfCont_Aopt_toa}A-optimal design.]{\begin{centering}
\includegraphics[scale=0.25]{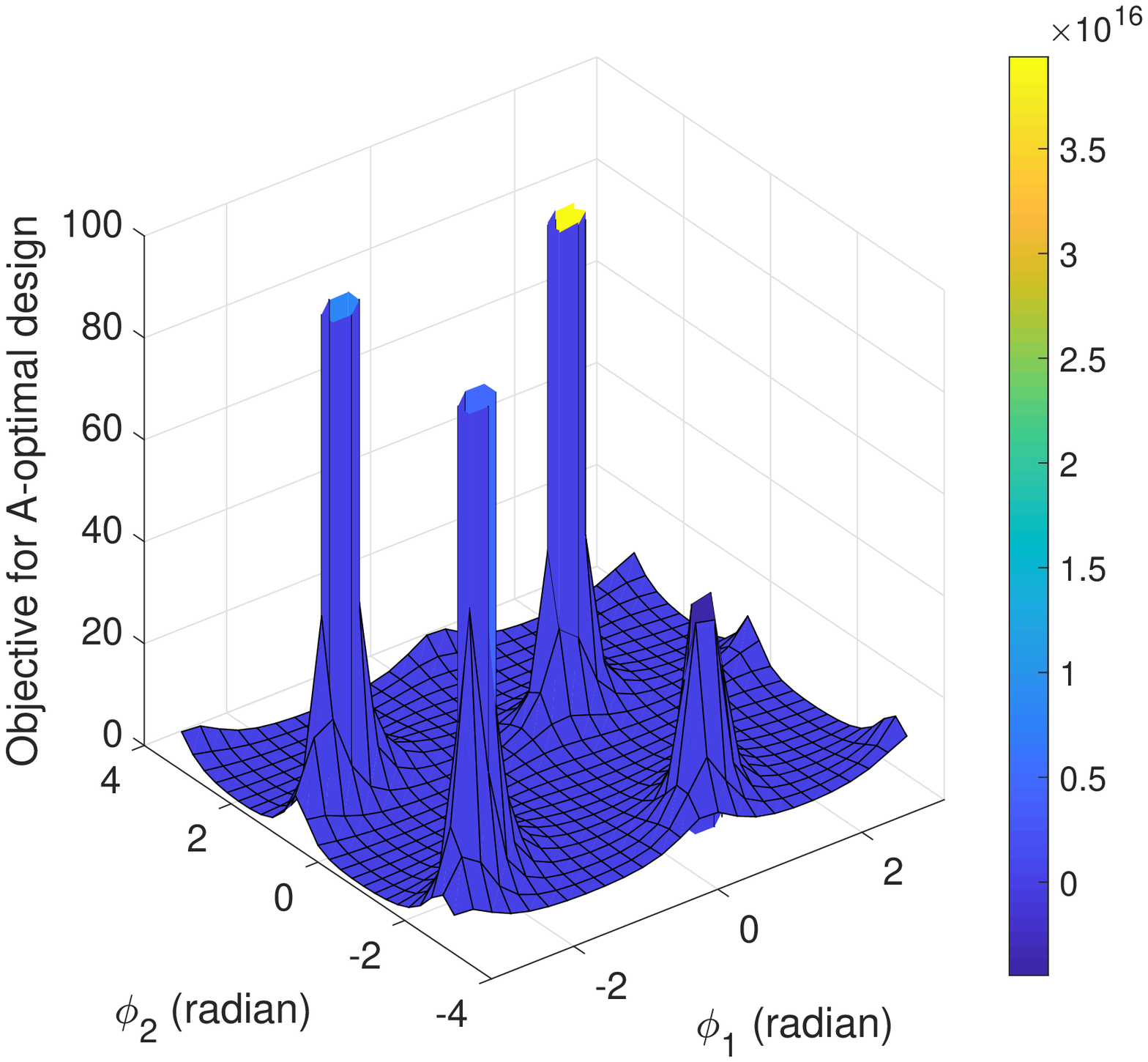}\includegraphics[scale=0.25]{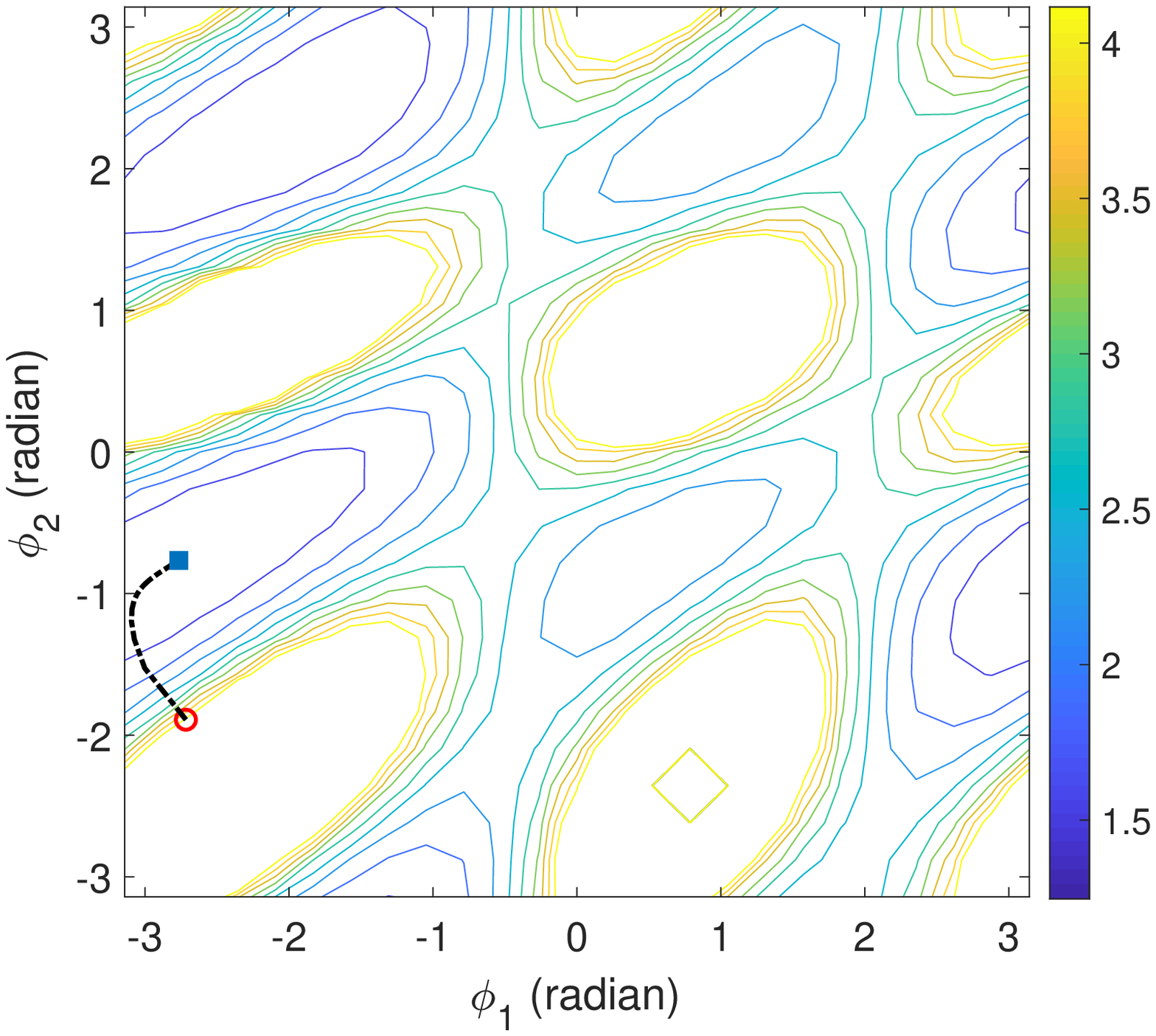}
\par\end{centering}
}
\par\end{centering}
\begin{centering}
\subfloat[\label{fig:SurfCont_Dopt_toa}D-optimal design]{\begin{centering}
\includegraphics[scale=0.25]{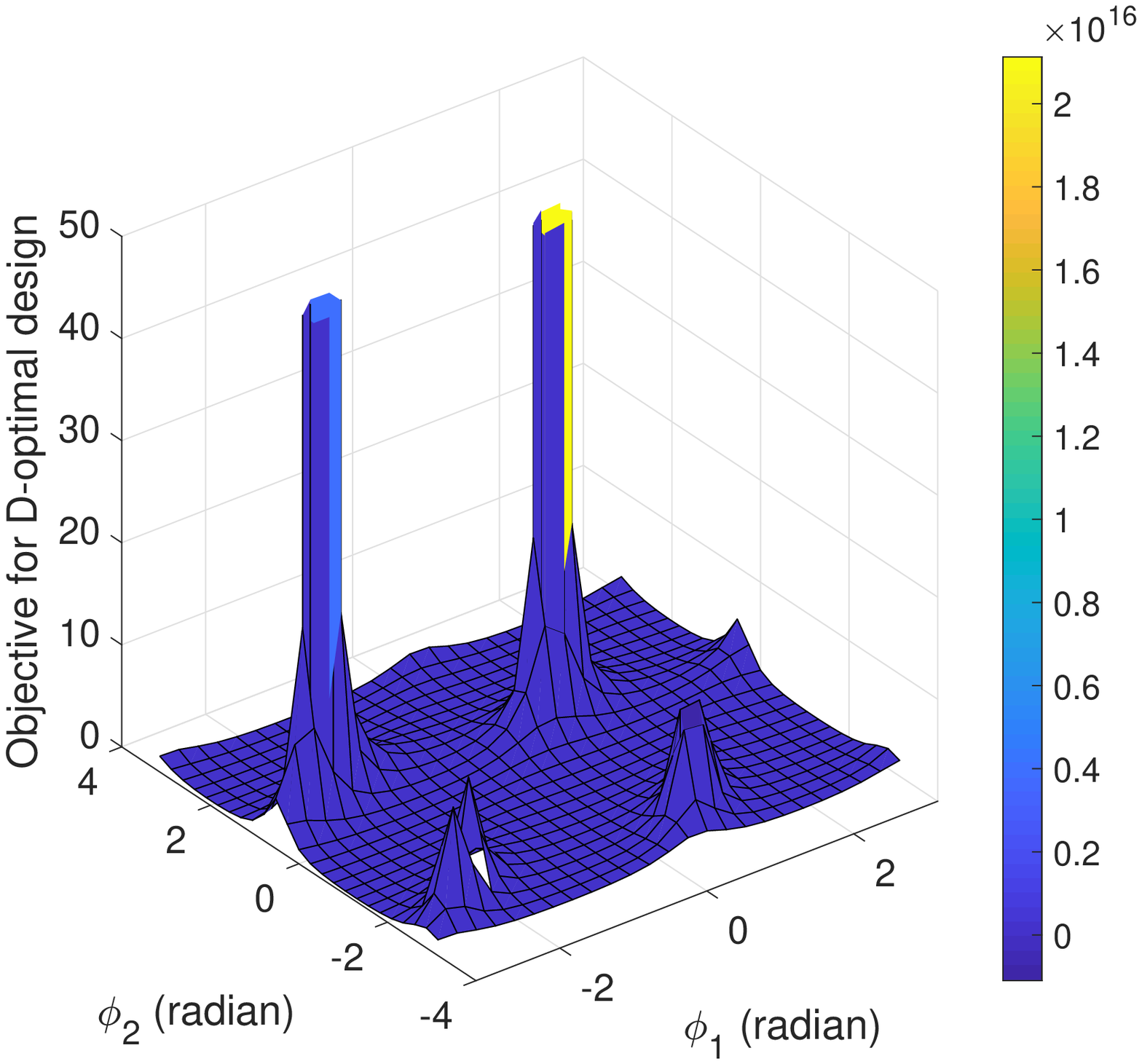}\includegraphics[scale=0.25]{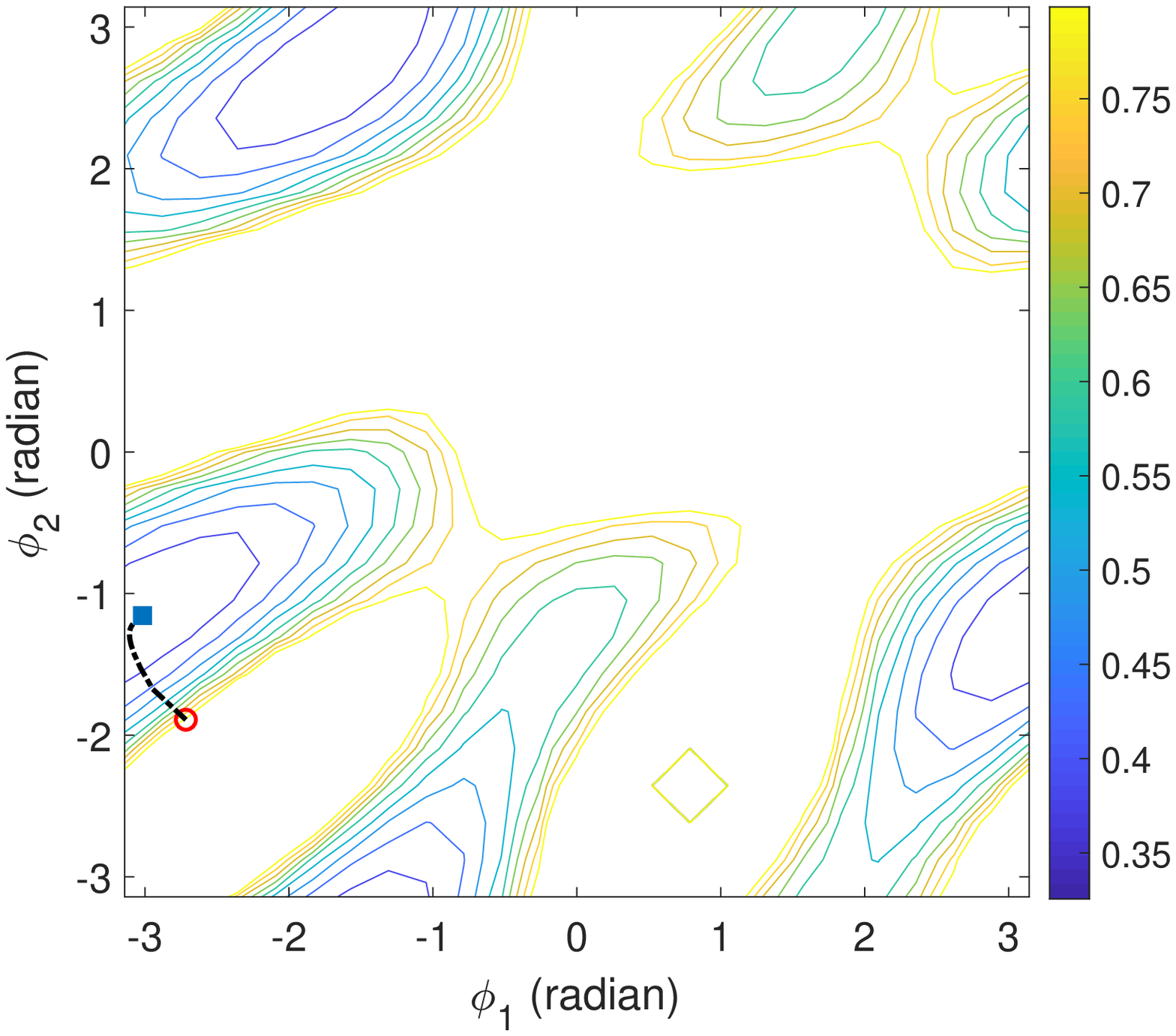}
\par\end{centering}
}
\par\end{centering}
\begin{centering}
\subfloat[\label{fig:SurfCont_Eopt_toa}E-optimal design.]{\begin{centering}
\includegraphics[scale=0.25]{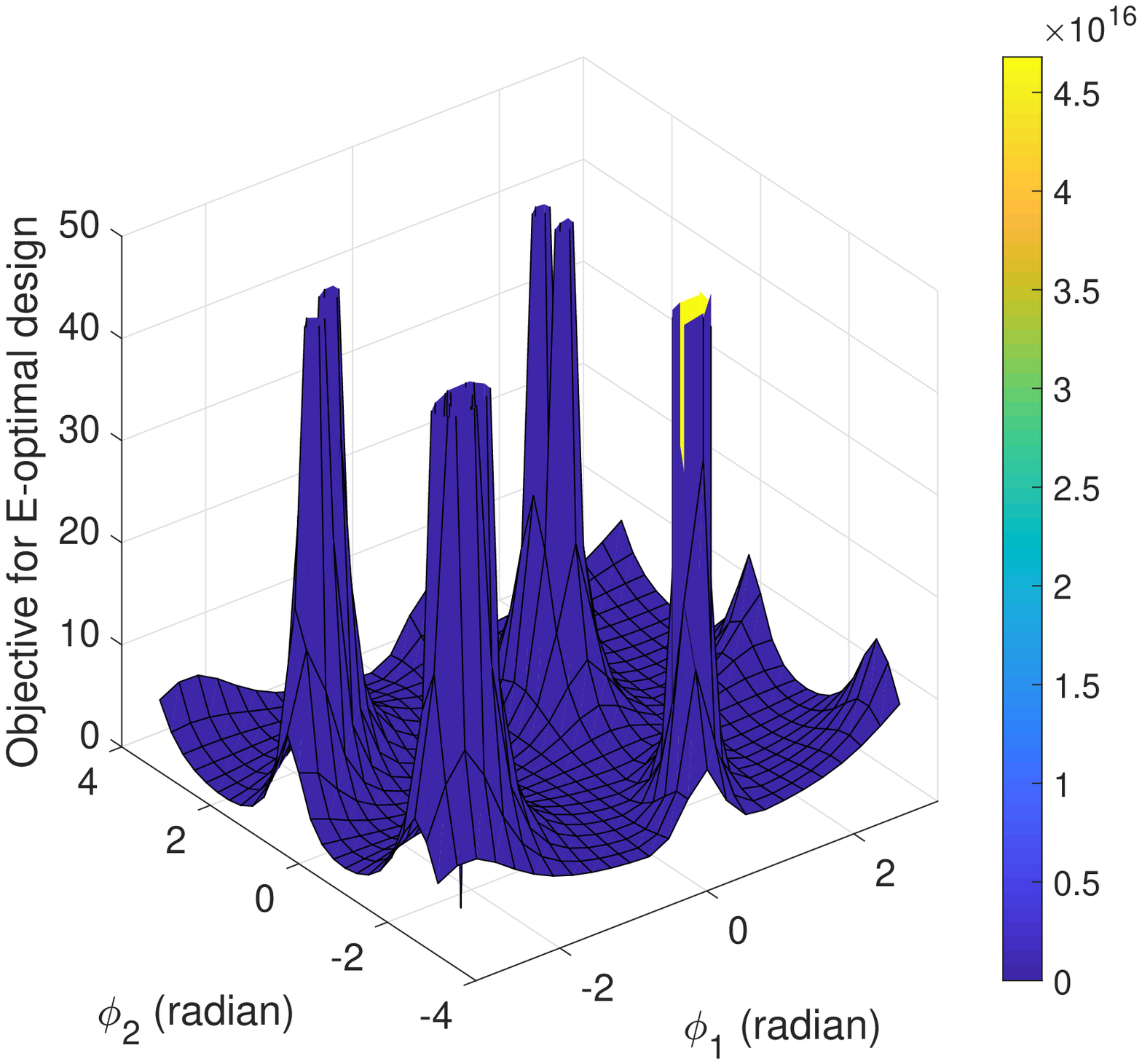}\includegraphics[scale=0.25]{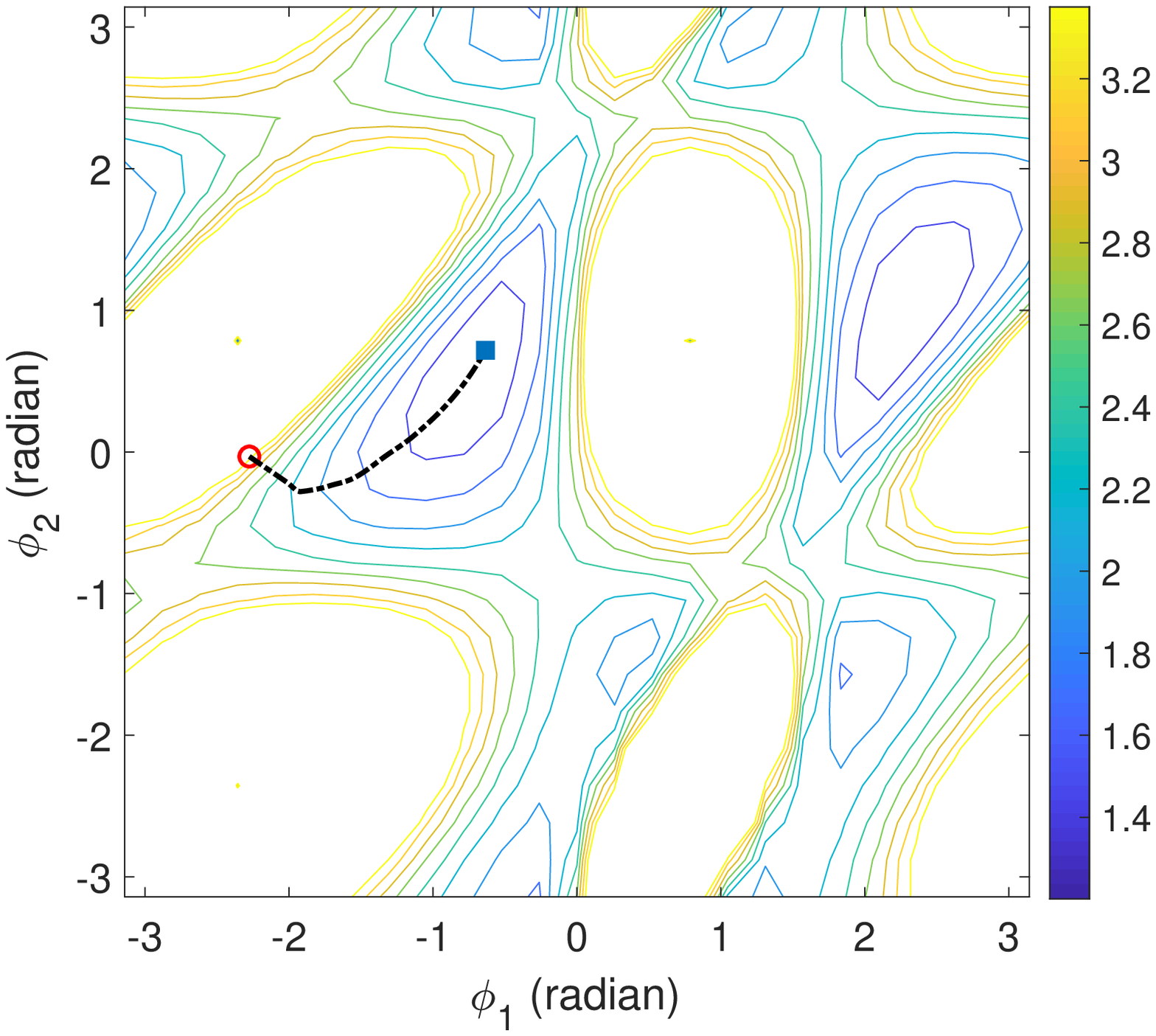}
\par\end{centering}
}
\par\end{centering}
\caption{\label{fig:Optimization-trajectory-for}Optimization trajectory for
2D TOA with some fixed sensor positions. Left: objective function
shape; right: contour plot. Red circle: initial value, blue square:
final optimal point.}
\end{figure}

In Figure \eqref{fig:Optimization-trajectory-for}, we diagrammatically
show how the proposed algorithmic framework finds the optimum of the
design objective. We consider a total of three sensors in 2D space
(i.e. $m=3,n=2$) and assume the sensors to be on unit circle with
the third sensor fixed at $\mathbf{r}_{3}=[\frac{1}{\sqrt{2}},\frac{1}{\sqrt{2}}]^{T}$.
As earlier target is coarsely located at the origin and we want to
determine the optimal placement of the remaining two sensors on the
unit circle (defined by their corresponding azimuth angles $\phi_{1}$
and $\phi_{2}$) to obtain maximum localization performance. In figure
\ref{fig:SurfCont_Aopt_toa}, \ref{fig:SurfCont_Dopt_toa} and \ref{fig:SurfCont_Eopt_toa},
the objective value has been plotted with $\phi_{1}$ and $\phi_{2}$
as surface plot and the path taken by the proposed algorithm to reach
the optimal solution is shown on the corresponding contour plots for
the A-, D- and E-optimal designs, respectively.

\subsection{TDOA Based Source Localization}

In this subsection, we perform simulations for the optimal sensor
placements for the TDOA model with $m=6$, $n=3$. We assume that
the first sensor is set as a reference for measuring TDOA's, so we
will have $m-1$ TDOA measurements to localize the target. The covariance
matrix $\mathbb{E}\left[\mathbf{n}\mathbf{n}^{T}\right]$ is associated
with the error in estimation of sensor-target range and assumed as
diagonal matrix with $\mathbb{E}\left[\mathbf{n}\mathbf{n}^{T}\right]=\mathrm{diag}\left(0.18,0.02,0.46,0.72,0.42,0.49\right)$.
Consequently, the noise covariance matrix associated with the TDOA
measurements is given by

\begin{equation}
\footnotesize{\mathbf{R}_{tdoa}=\mathbf{K}\mathbb{E}\left[\mathbf{n}\mathbf{n}^{T}\right]\mathbf{K}^{T}=\left[\begin{array}{ccccc}
0.20 & 0.18 & 0.18 & 0.18 & 0.18\\
0.18 & 0.64 & 0.18 & 0.18 & 0.18\\
0.18 & 0.18 & 0.91 & 0.18 & 0.18\\
0.18 & 0.18 & 0.18 & 0.60 & 0.18\\
0.18 & 0.18 & 0.18 & 0.18 & 0.67
\end{array}\right].}
\end{equation}
Similar to the TOA case, we initialize the proposed algorithm with
the uniform placement as given in \eqref{eq:SimRslt_e2}.

\begin{figure}[t]
\begin{centering}
\subfloat[A-optimal design.]{\begin{centering}
\includegraphics[scale=0.25]{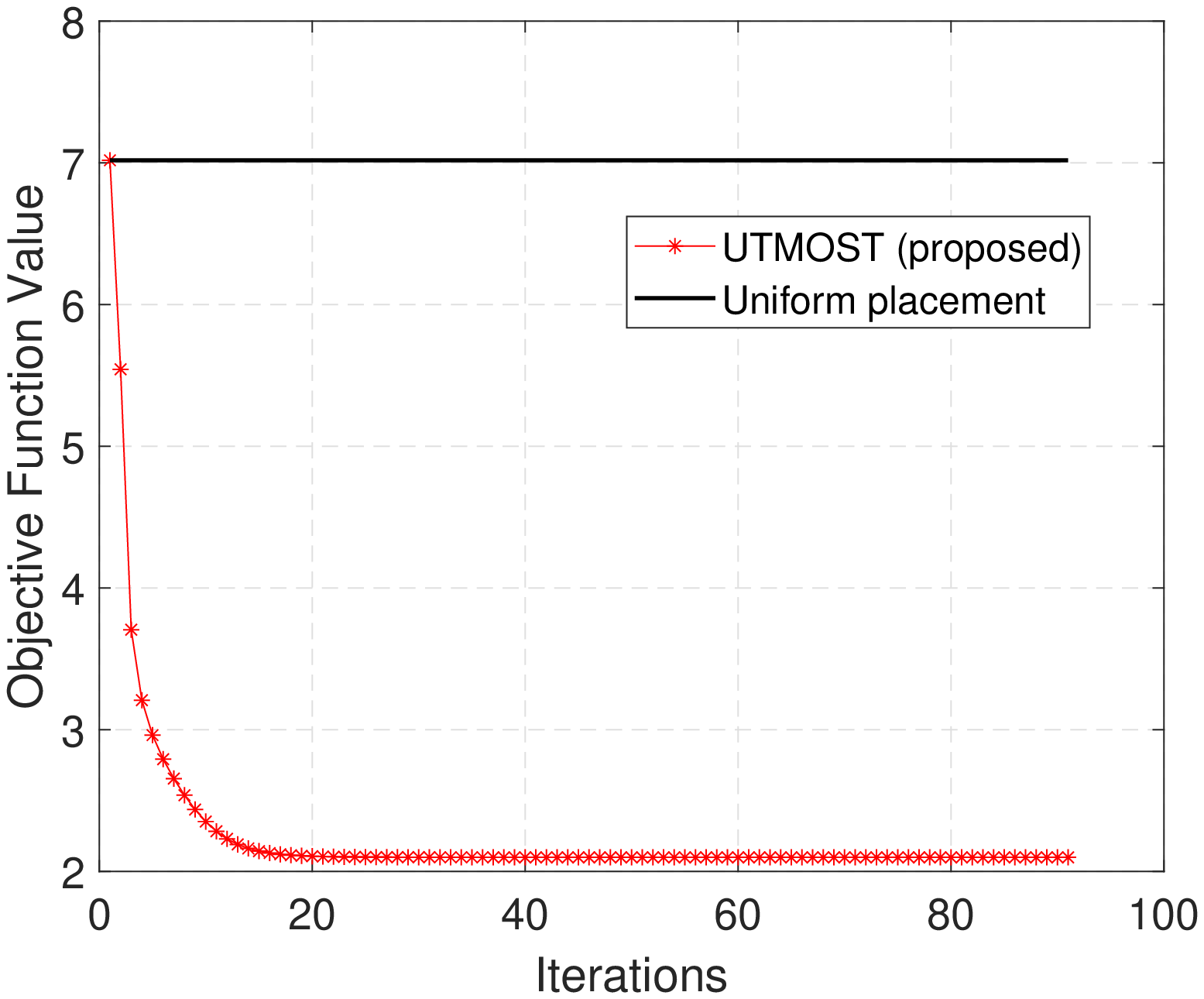}\includegraphics[scale=0.25]{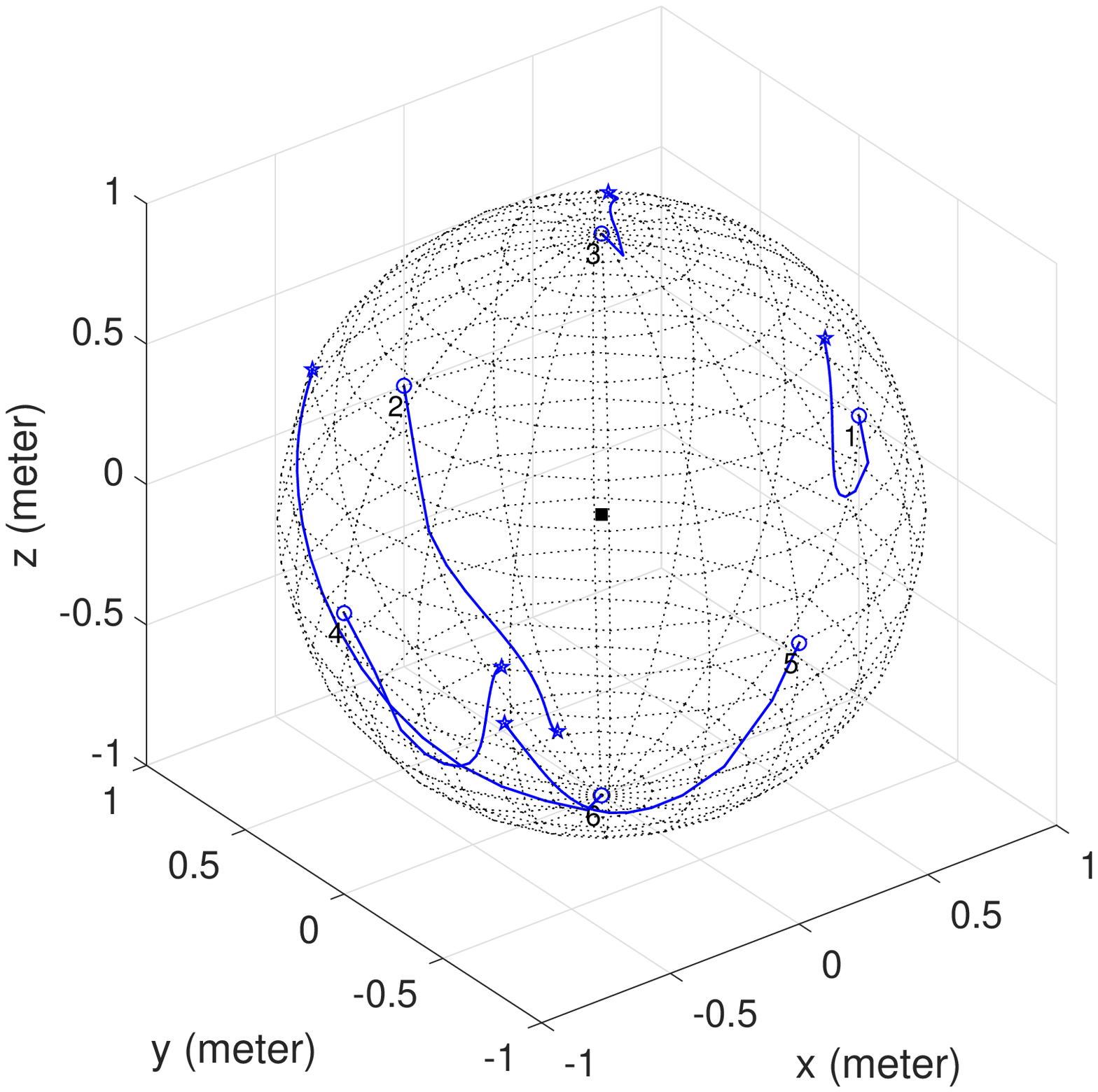}
\par\end{centering}
}
\par\end{centering}
\begin{centering}
\subfloat[D-optimal design]{\begin{centering}
\includegraphics[scale=0.25]{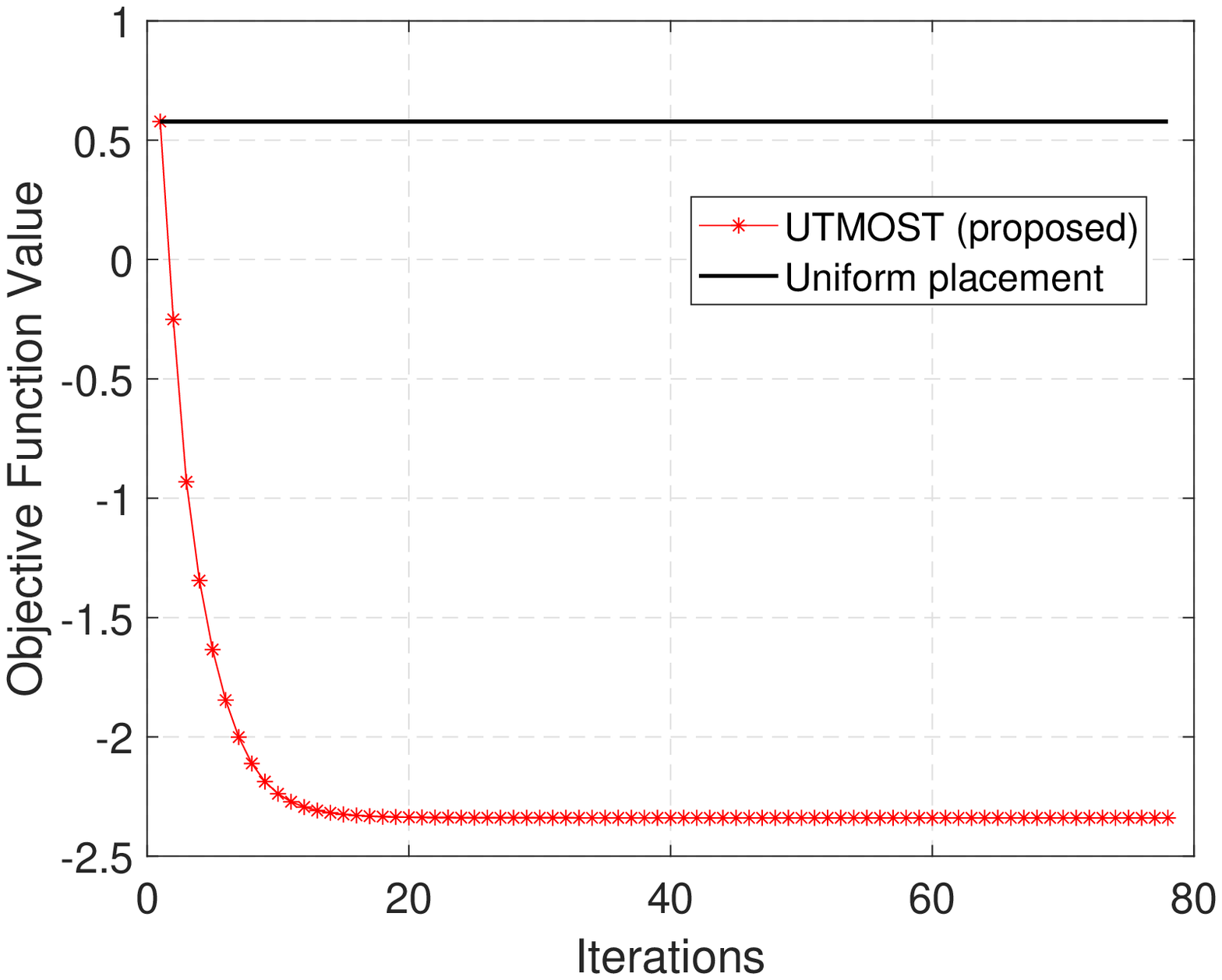}\includegraphics[scale=0.25]{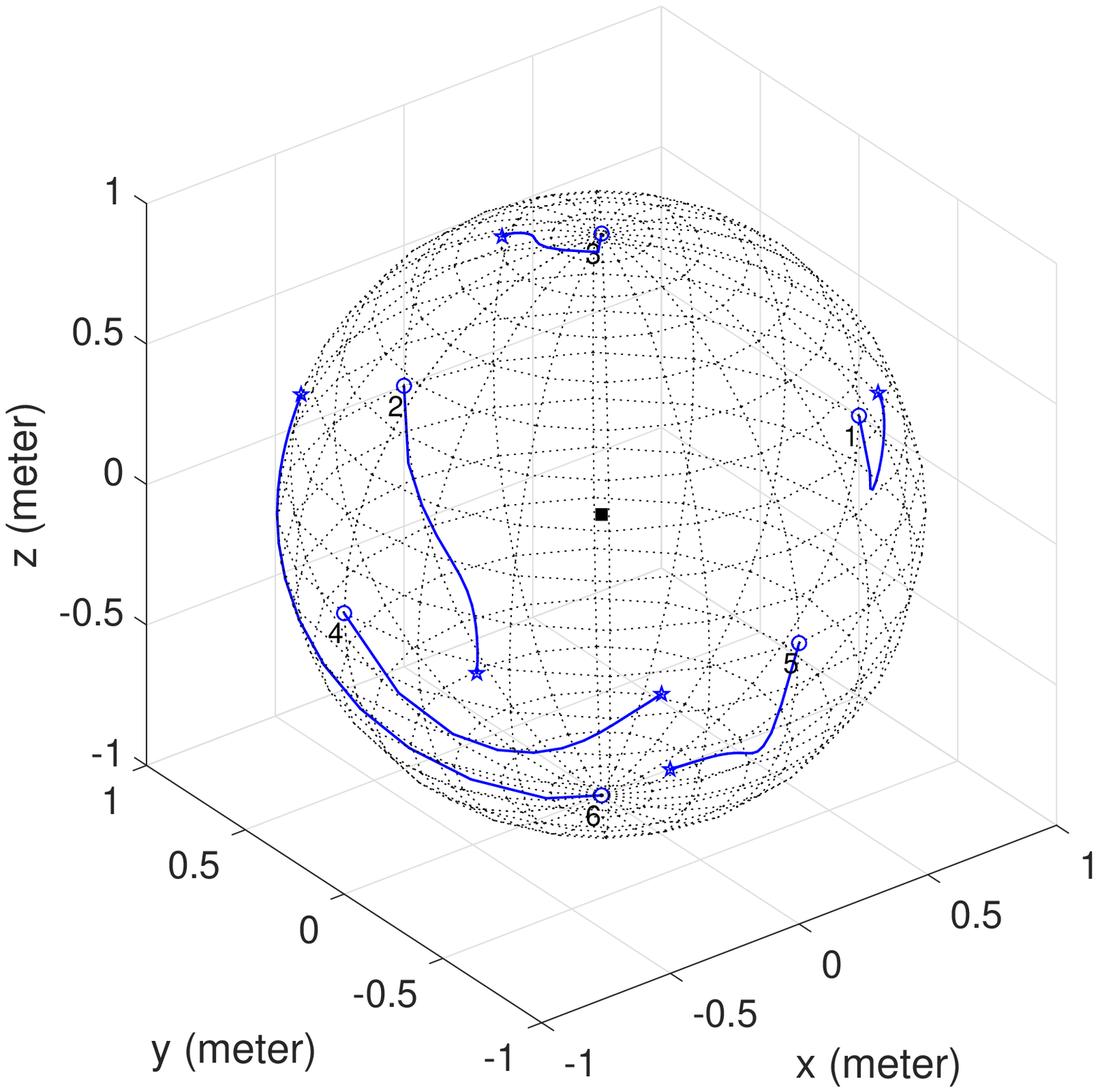}
\par\end{centering}
}
\par\end{centering}
\begin{centering}
\subfloat[E-optimal design.]{\begin{centering}
\includegraphics[scale=0.25]{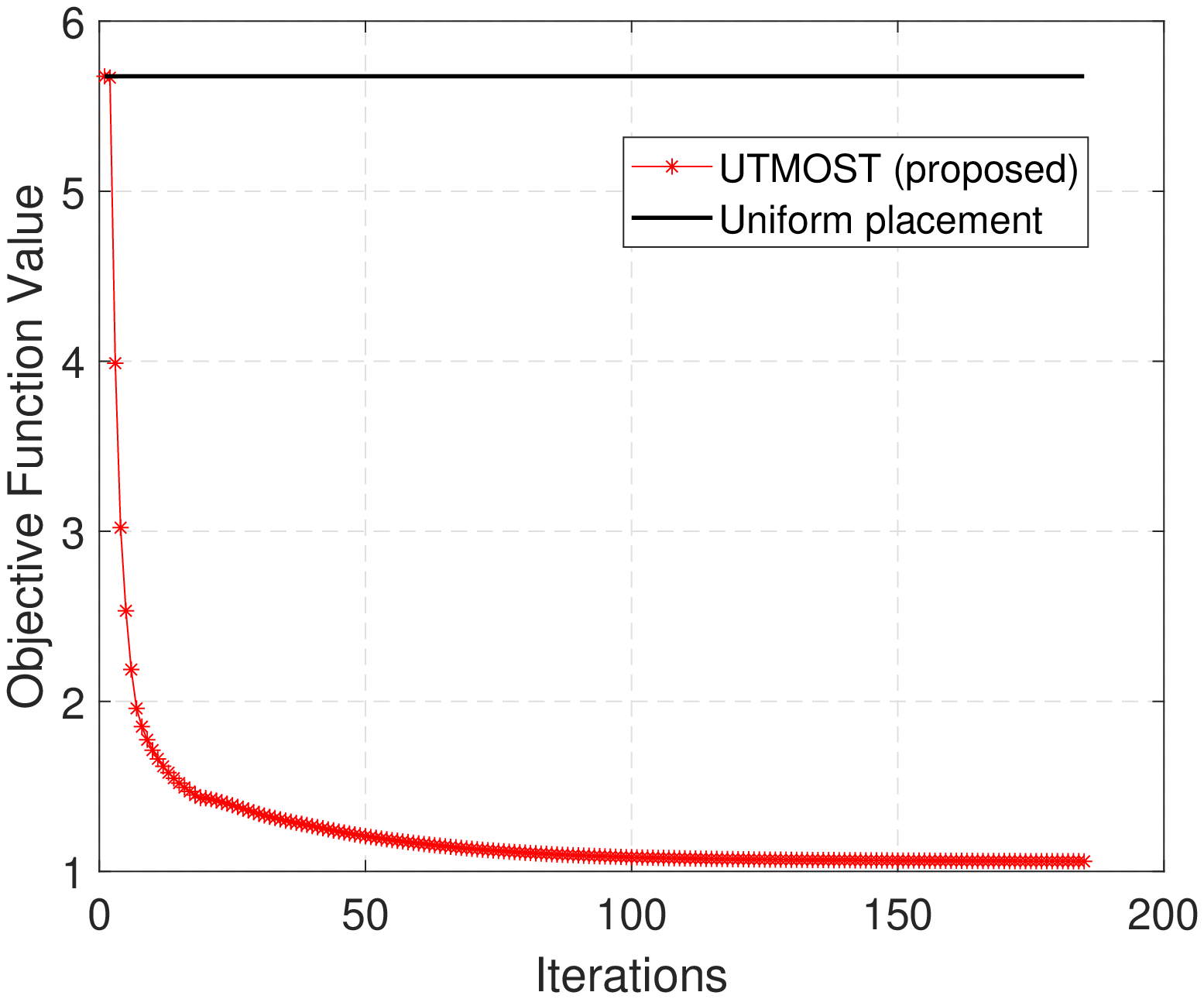}\includegraphics[scale=0.25]{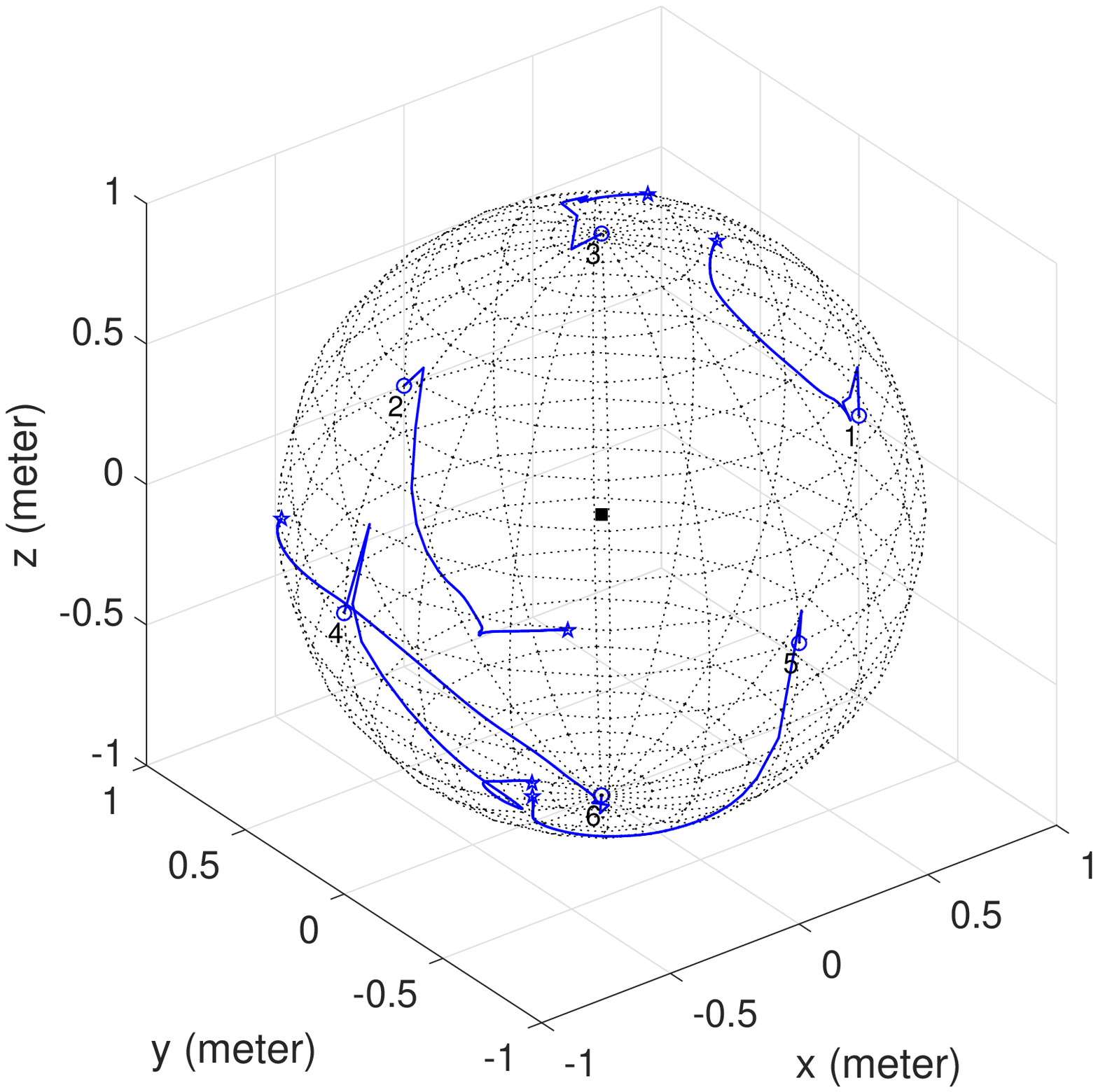}
\par\end{centering}
}
\par\end{centering}
\caption{\label{fig:ConvPlot_tdoa}Convergence plots and corresponding sensor
placements for the TDOA based model under the correlated measurement
noise. Left: convergence plots; right: 3D view of sensor placement.
Black square: target; blue circle: initial position; blue pentagram:
end position.}
\end{figure}

Figure \ref{fig:ConvPlot_tdoa} demonstrates the convergence plots
and the corresponding sensor placements. The difference of the objective
values between the uniform and proposed (after convergence) placements
shows the $70-80\%$ improvement in localization accuracy obtained
by the proposed algorithm when the measurement noise is correlated.
In the right column of Figure \ref{fig:ConvPlot_tdoa}, we can see
how the sensors move from the initial uniform placement to the final
optimal configuration (obtained by the proposed algorithm) for the
three optimal designs, respectively. 

\subsection{RSS Based Source Localization}

In this subsection, we perform simulations for the RSS model. Recall
that both AOA and RSS has the same formulations and assumptions (see
Remark \ref{rem:Remark 5} and expression \eqref{eq:3_CRLB}), the
simulations of AOA will demonstrate the same patterns as RSS and thus
be omitted. Unlike the TOA and TDOA cases, in the RSS-based localization,
the optimal placement also depends on the sensor-target range. Assuming
that the sensor-target range for each sensor is given (or roughly
known), we are interested in determining the optimal configuration.
Let $m=6$, $n=3$, and the sensor-target range $d_{i}$ (in meter)
and the noise covariance matrix $\mathbf{R}_{rss}$ are set as, respectively,
\begin{equation}
\left[d_{1},d_{2},d_{3},d_{4},d_{5},d_{6}\right]=\left[50,100,150,200,250,300\right],
\end{equation}
and

\begin{equation}
\mathbf{R}_{rss}=\left[\begin{array}{cccccc}
4.88 & 3.07 & -1.73 & 1.90 & 2.63 & -1.61\\
3.07 & 11.72 & -3.51 & 4.48 & 3.95 & 0.24\\
-1.73 & -3.51 & 21.82 & -1.20 & 0.49 & -4.74\\
1.90 & 4.48 & -1.20 & 3.63 & 3.71 & 1.00\\
2.63 & 3.95 & 0.49 & 3.71 & 8.45 & 0.56\\
-1.61 & 0.24 & -4.74 & 1.00 & 0.56 & 4.22
\end{array}\right].
\end{equation}
Therefore, by taking $\mathbf{D}=\mathrm{diag}\left(d_{1},\ldots,d_{6}\right)^{-1}$,
we can use the proposed algorithm to compute the optimal configuration. 

\begin{figure}[!t]
\begin{centering}
\subfloat[A-optimal design.]{\begin{centering}
\includegraphics[scale=0.25]{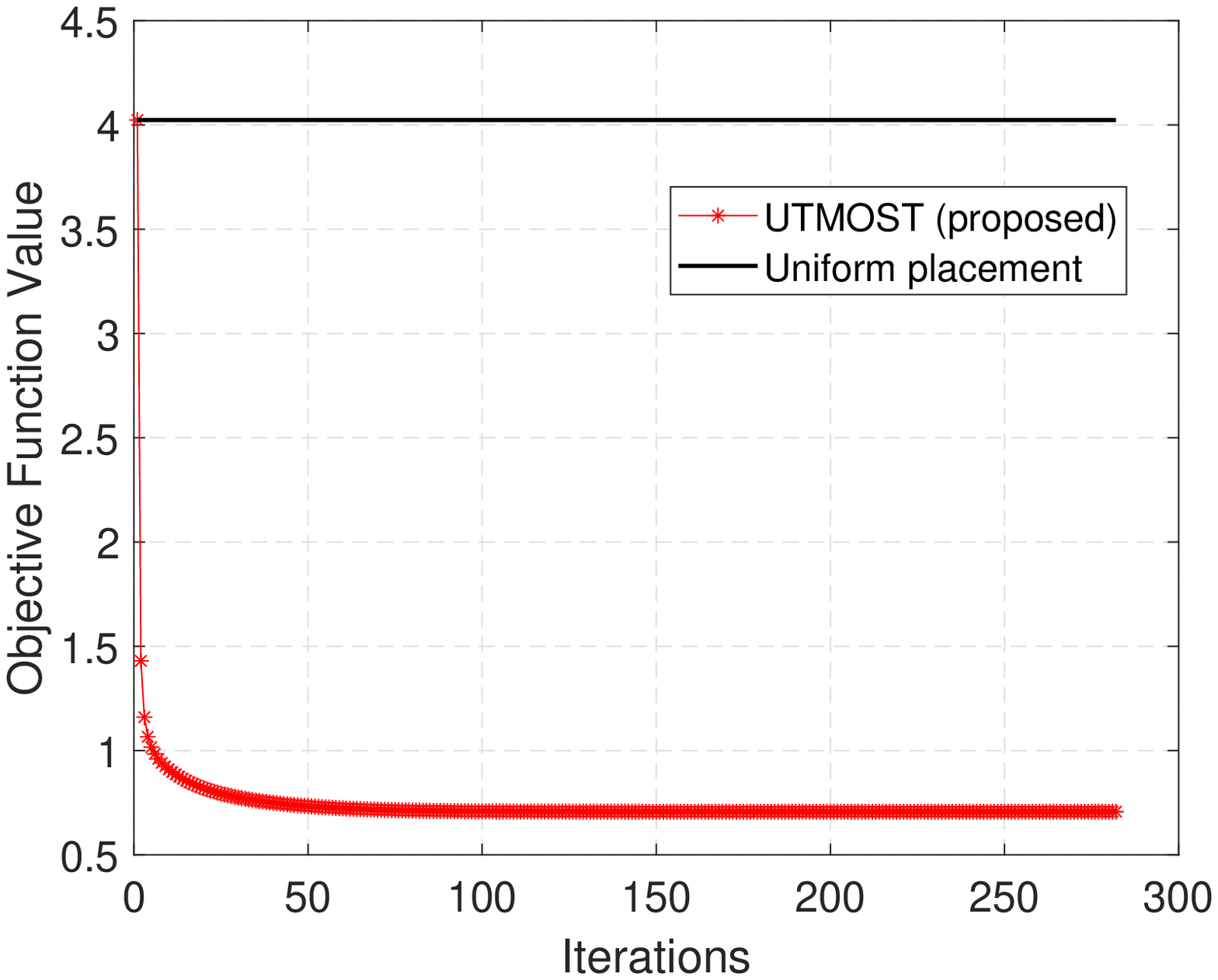}\includegraphics[scale=0.25]{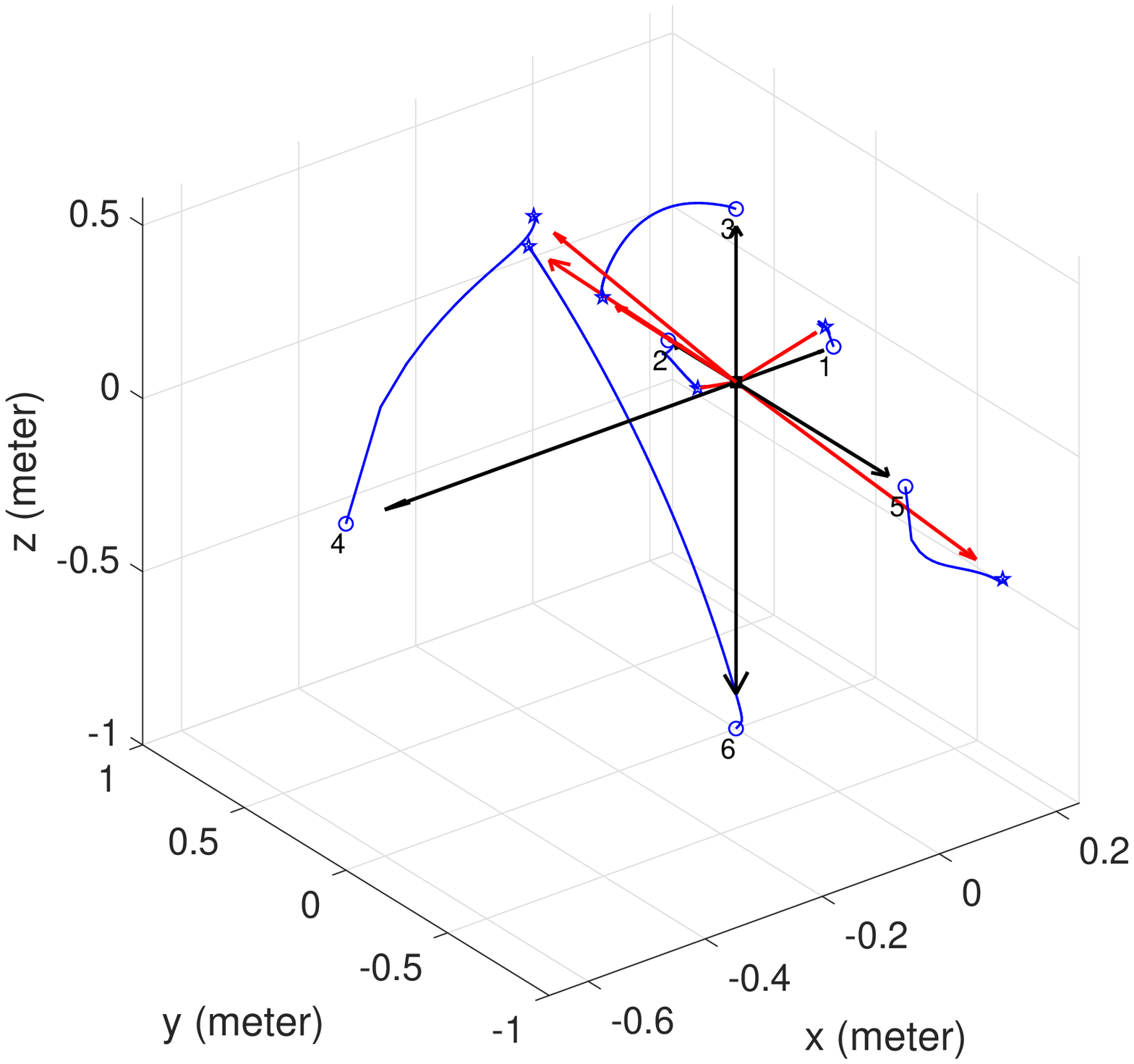}
\par\end{centering}
}
\par\end{centering}
\begin{centering}
\subfloat[D-optimal design]{\begin{centering}
\includegraphics[scale=0.25]{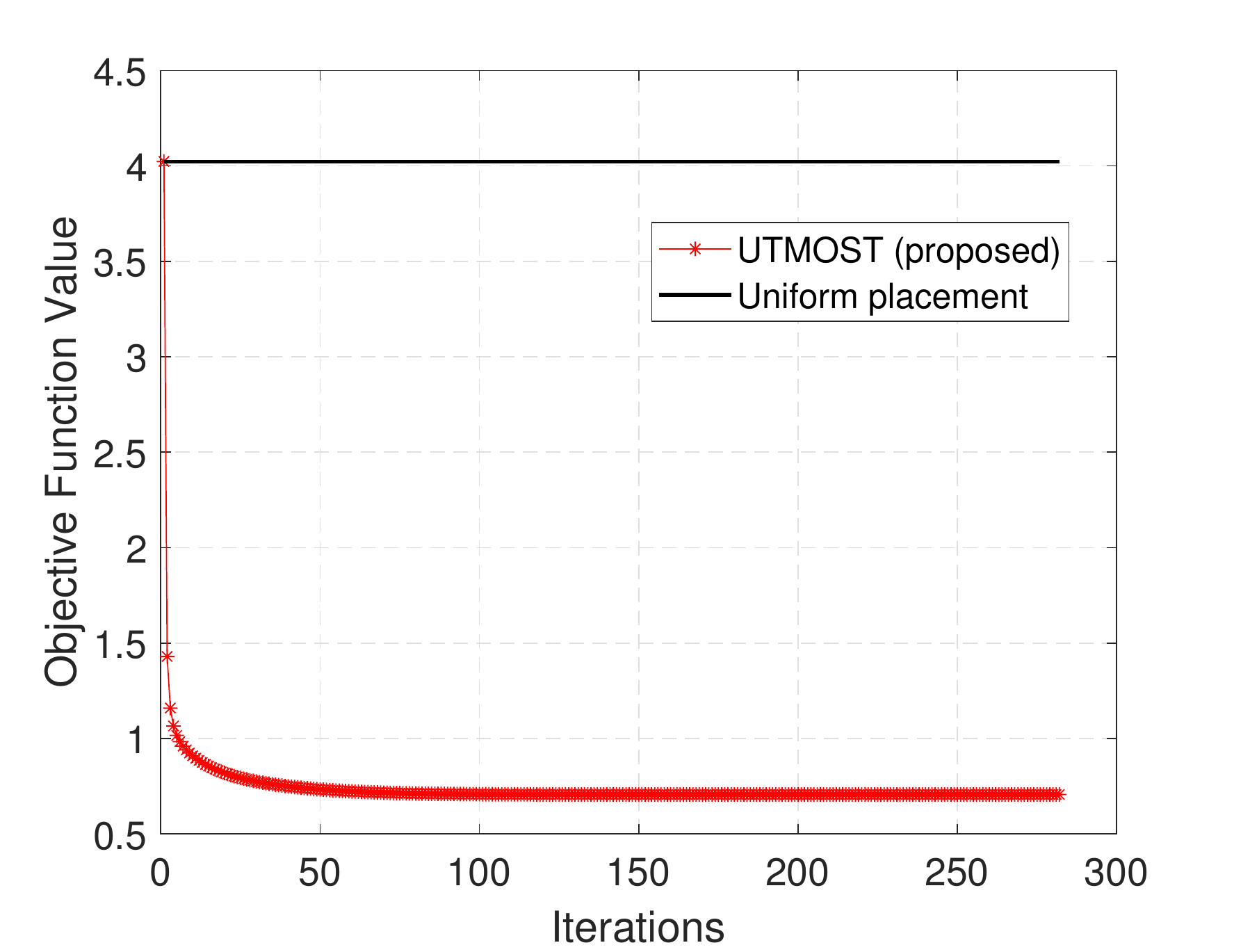}\includegraphics[scale=0.25]{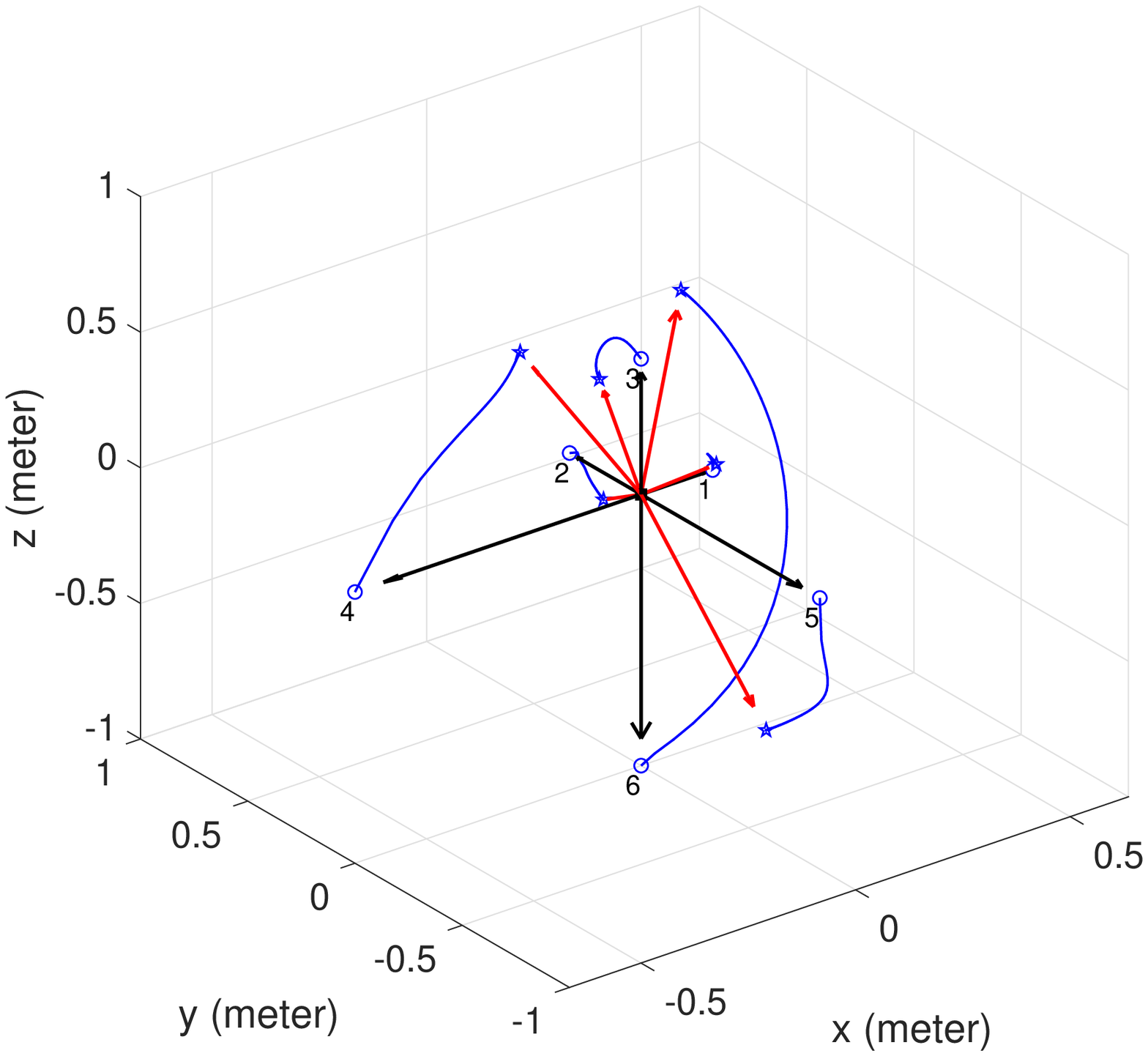}
\par\end{centering}
}
\par\end{centering}
\begin{centering}
\subfloat[E-optimal design.]{\begin{centering}
\includegraphics[scale=0.25]{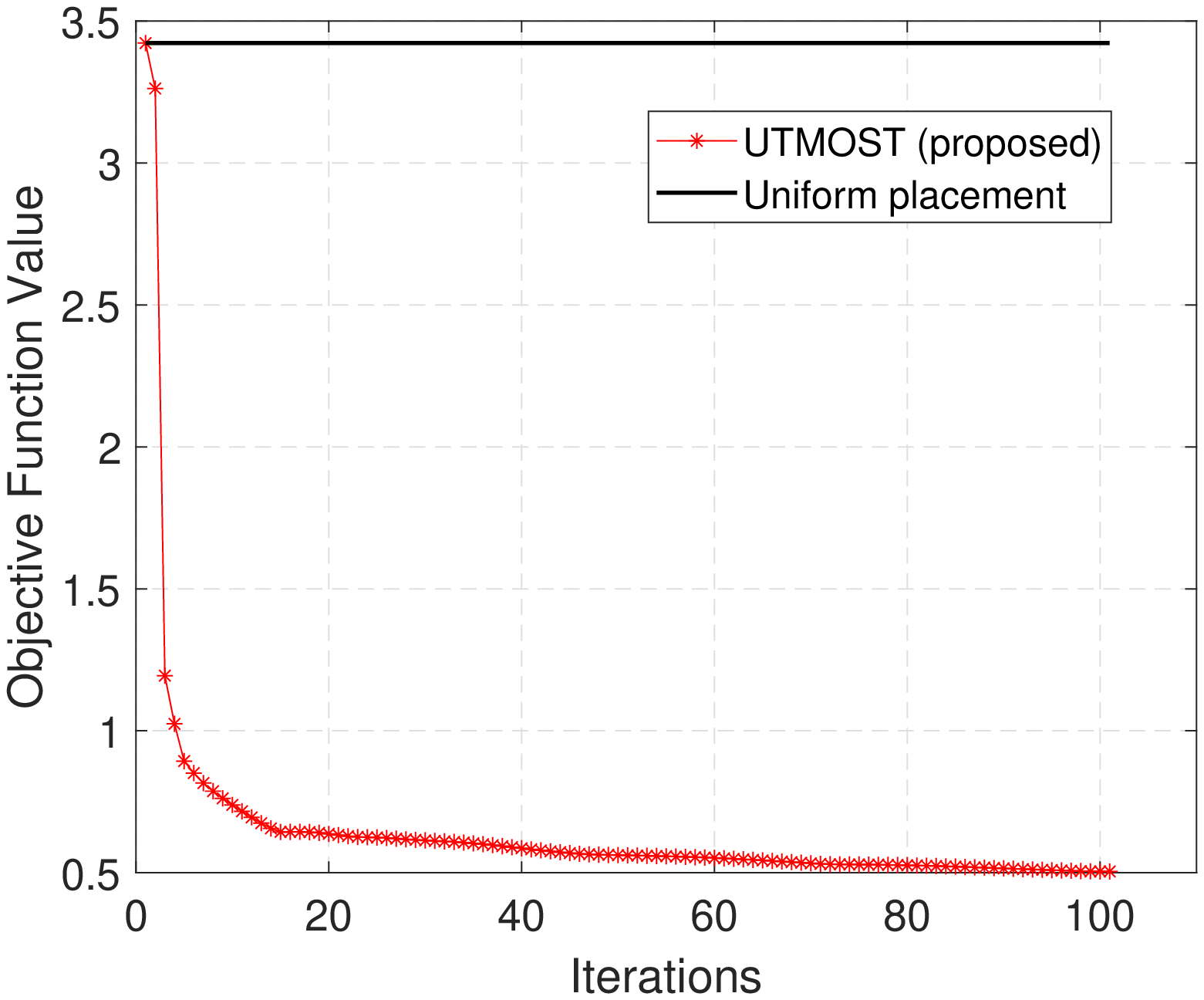}\includegraphics[scale=0.25]{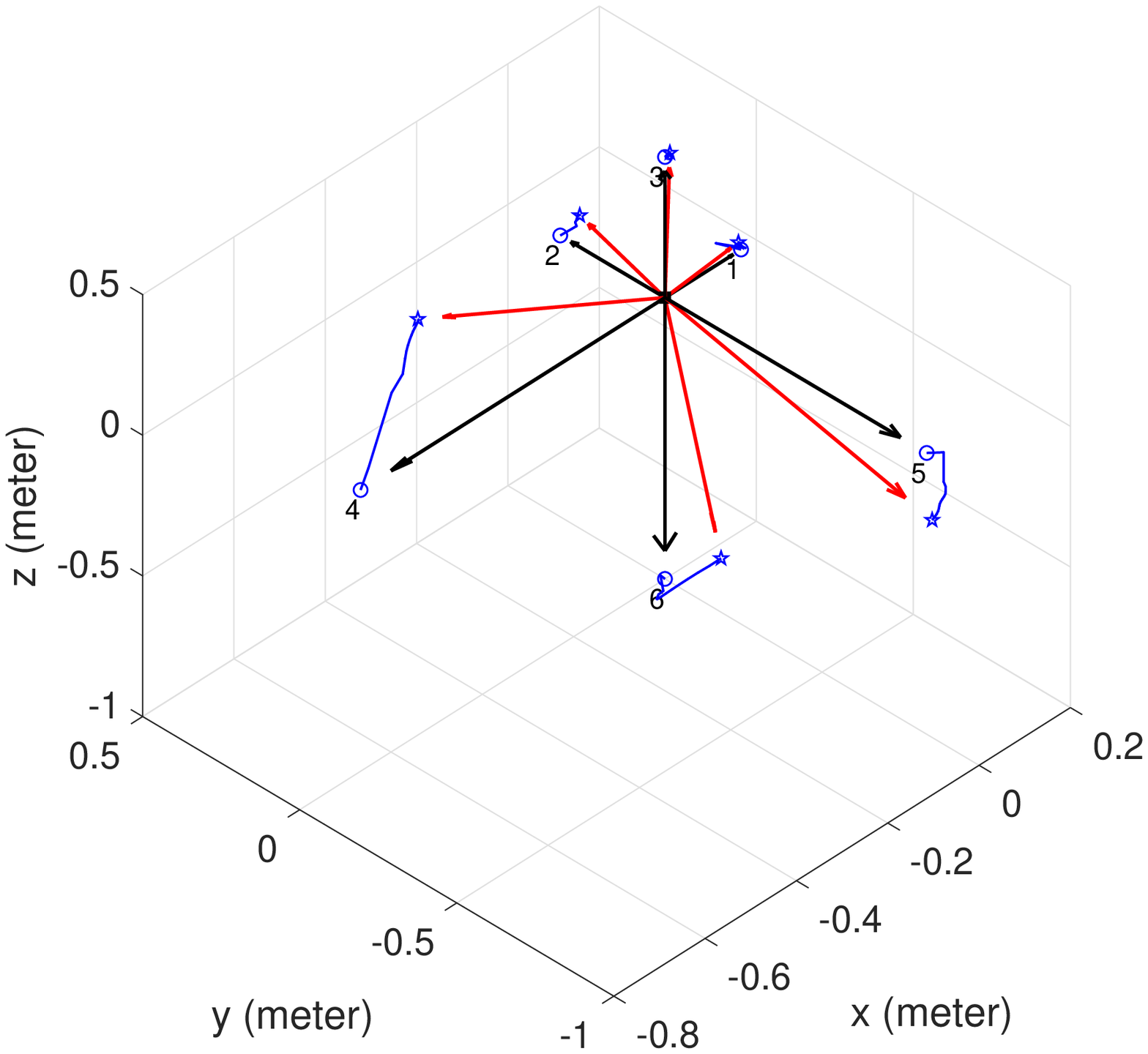}
\par\end{centering}
}
\par\end{centering}
\caption{\label{fig:ConvPlot_rss}Convergence plots and corresponding sensor
placements for the RSS based model under the correlated measurement
noise. Left: convergence plots; right: 3D view of sensor placement.
Black square: target; blue circle: initial position; blue pentagram:
end position.}
\end{figure}

Figure \ref{fig:ConvPlot_rss} demonstrates the convergence plots
and the sensor placements of the proposed method with reference to
the uniform placement (the sensors are uniformly placed w.r.t. the
target), where the $i$-th sensor is shown to be at distance $\frac{d_{i}}{\max\left\{ d_{i}\right\} }$
from the target. in the 3D plot. The difference in objective values
of the uniform placement and proposed method (after convergence) shows
the $80-85\%$ improvement (in terms of the design criteria) in localization
accuracy obtained by the proposed algorithm when the measurement noise
is correlated. 

Next, we illustrate the impact of having imperfect knowledge of range
matrix $\mathbf{D}$. To demonstrate this we perform a simulation
to compute the MLE (similar to table-\ref{tab:MLE-performance.} for
the TOA model) for the RSS model, in which we evaluate the MSEs of
the MLE method for the cases when $\mathbf{D}$ is perfectly known
and $\mathbf{D}$ is imperfectly known. The true target position is
taken as $\left(0.1,-0.3\right)$ and in case of ``Coarse $\mathbf{D}$''
we assumed that coarse target position is at $\left(0.0,0.0\right)$
and calculate the range matrix. The MSEs of the MLE method (solved
using grid search and additional Gauss-Newton step) for the case of
optimal sensor placement done with the perfect knowledge of $\mathbf{D}$
and coarsely known $\mathbf{D}$ are shown in table \ref{tab:MLE-performance RSS}
here. It can be seen from table \ref{tab:MLE-performance RSS}, the
MSE in the case of coarsely known $\mathbf{D}$ is higher than the
case of perfectly known $\mathbf{D}$, which is as expected.

\begin{table}[!t]
\caption{Comparison of the MLE performance for RSS model.\label{tab:MLE-performance RSS}}

\centering{}%
\begin{tabular}{ccccc}
\toprule 
 & \multicolumn{2}{c}{$\mathbf{D}$ perfectly known} & \multicolumn{2}{c}{$\mathbf{D}$ coarsely known}\tabularnewline
\midrule 
No. of sensors & MSE ($\text{m}^{2}$) & Bias ($\text{m}$) & MSE ($\text{m}^{2}$) & Bias ($\text{m}$)\tabularnewline
\midrule
\midrule 
$m=3$ & $0.5540$ & $0.0872$ & $0.5627$ & $0.0972$\tabularnewline
\midrule 
$m=4$ & $0.2797$ & $0.2467$ & $0.3020$ & $0.2440$\tabularnewline
\midrule 
$m=5$ & $0.3073$ & $0.0846$ & $0.3116$ & $0.0865$\tabularnewline
\bottomrule
\end{tabular}
\end{table}

\section{Conclusions\label{sec:Conclusion}}

In this paper, we have unified the three TOA, TDOA, AOA and RSS based
sensor placement case in a generalized problem formulation based on
the CRLB-related metric. For this general problem, we have developed
a unified optimization approach named UTMOST based on the ADMM and
MM techniques. Within in this framework, we can handle the sensor
placement for all the TOA, TDOA, and RSS based source localization
methods by specifying the system parameters. For each localization
model, this framework can be adapted with slight modifications to
design the sensor placement under all the A-, D- and E-optimality
criteria. Through the numerical simulations, we have demonstrated
the versatility of the unified approach by considering various placement
scenarios and also the improvement of localization accuracy brought
by the optimal configuration of sensors.

\appendices{}

\appendix{}

\subsection{Proof of Lemma \ref{lem:Lemma4}\label{subsec:AppenA}}
\begin{IEEEproof}
The objective function of problem \eqref{eq:subproblem_Y} can be
written as

\begin{equation}
\begin{aligned} & f\left(\left(\mathbf{Y}^{T}\mathbf{Y}\right)^{-1}\right)+\frac{\rho}{2}\mathrm{Tr}\left(\mathbf{Y}^{T}\mathbf{R}\mathbf{Y}\right)-\mathrm{Tr}\left(\mathbf{E}_{k}^{T}\mathbf{Y}\right)\\
= & f\left(\left(\mathbf{Y}^{T}\mathbf{Y}\right)^{-1}\right)-\mathrm{Tr}\left(\mathbf{E}_{k}^{T}\mathbf{Y}\right)\\
 & \frac{\rho}{2}\mathrm{Tr}\left(\mathbf{Y}^{T}\mathbf{R}\mathbf{Y}-\lambda_{m}\left(\mathbf{R}\right)\mathbf{Y}^{T}\mathbf{Y}+\lambda_{m}\left(\mathbf{R}\right)\mathbf{Y}^{T}\mathbf{Y}\right)\\
= & f\left(\left(\mathbf{Y}^{T}\mathbf{Y}\right)^{-1}\right)+\frac{\rho}{2}\mathrm{Tr}\left(\mathbf{Y}^{T}\widetilde{\mathbf{R}}\mathbf{Y}\right)\\
 & \frac{\rho}{2}\lambda_{m}\left(\mathbf{R}\right)\mathrm{Tr}\left(\mathbf{Y}^{T}\mathbf{Y}\right)-\mathrm{Tr}\left(\mathbf{E}_{k}^{T}\mathbf{Y}\right),
\end{aligned}
\label{eq:sec3_e6}
\end{equation}
where $\widetilde{\mathbf{R}}\triangleq\mathbf{R}-\lambda_{m}\left(\mathbf{R}\right)\mathbf{I}_{m}$.
Since $\widetilde{\mathbf{R}}\preceq0$, $\mathrm{Tr}\left(\mathbf{Y}^{T}\widetilde{\mathbf{R}}\mathbf{Y}\right)$
is concave of $\mathbf{Y}$. Its first order Taylor expansion satisfies

\begin{equation}
\mathrm{Tr}\left(\mathbf{Y}^{T}\widetilde{\mathbf{R}}\mathbf{Y}\right)\leq2\mathrm{Tr}\left(\mathbf{Y}_{\tau}^{T}\widetilde{\mathbf{R}}\mathbf{Y}\right)-\mathrm{Tr}\left(\mathbf{Y}_{\tau}^{T}\widetilde{\mathbf{R}}\mathbf{Y}_{\tau}\right)\label{eq:sec3_e7}
\end{equation}
for all $\mathbf{Y}$ with the equality achieved at $\mathbf{Y}=\mathbf{Y}_{\tau}$. 

Applying \eqref{eq:sec3_e7} on \eqref{eq:sec3_e6}, we have the upper
bound given by 
\begin{equation}
\begin{aligned} & g_{L}\left(\mathbf{Y}\right)\\
= & f\left(\left(\mathbf{Y}^{T}\mathbf{Y}\right)^{-1}\right)+\frac{\rho}{2}\lambda_{m}\left(\mathbf{R}\right)\mathrm{Tr}\left(\mathbf{Y}^{T}\mathbf{Y}\right)-\mathrm{Tr}\left(\mathbf{E}_{k}^{T}\mathbf{Y}\right)\\
 & +\frac{\rho}{2}\left(2\mathrm{Tr}\left(\mathbf{Y}_{\tau}^{T}\widetilde{\mathbf{R}}\mathbf{Y}\right)-\mathrm{Tr}\left(\mathbf{Y}_{\tau}^{T}\widetilde{\mathbf{R}}\mathbf{Y}_{\tau}\right)\right)\\
= & f\left(\left(\mathbf{Y}^{T}\mathbf{Y}\right)^{-1}\right)+\frac{\rho}{2}\lambda_{m}\left(\mathbf{R}\right)\mathrm{Tr}\left(\mathbf{Y}^{T}\mathbf{Y}\right)\\
 & -\mathrm{Tr}\left(\left(\mathbf{E}_{k}-\rho\widetilde{\mathbf{R}}\mathbf{Y}_{\tau}\right)^{T}\mathbf{Y}\right)-\frac{\rho}{2}\mathrm{Tr}\left(\mathbf{Y}_{\tau}^{T}\widetilde{\mathbf{R}}\mathbf{Y}_{\tau}\right),
\end{aligned}
\end{equation}
which thereby completes the proof.
\end{IEEEproof}
\bibliographystyle{IEEEtran}
\bibliography{IEEEabrv,IEEEexample,ReferencesMSDF}

\end{document}